\documentclass[manuscript,nonacm,screen]{acmart}
\usepackage{blindtext, graphicx}
\usepackage{microtype}
\usepackage{mathtools}

\usepackage{algorithm}
\usepackage[noend]{algpseudocode}
\usepackage{url}
\usepackage{todonotes}
\usepackage{textcomp}
\usepackage{parskip}
\usepackage{pgfplots}
\usepackage{pgfplotstable}
\usepgfplotslibrary{statistics}
\pgfplotsset{compat=newest}
\usepgfplotslibrary{colorbrewer}
\pgfplotsset{cycle list/Dark2-8}
\usepackage{pgfplotstable}
\usepackage{booktabs}
\usepackage{tabularx}
\usepackage{multirow}
\usepackage{tablefootnote}
\usepackage{footmisc}
\usepackage{fancyvrb}
\usepackage{listings}
\usepackage{comment}
\usepackage{acronym}
\usepackage{bm, bbm}
\usepackage{diagbox}
\usepgfplotslibrary{groupplots}
\makeatletter

\pgfplotsset{
    discard if not A/.style 2 args={
        x filter/.append code={
            \edef\tempa{#1}
            \edef\tempaa{#2}
            \ifx\tempa\tempaa
            \else
                
            \fi
        }
    }}
\makeatother
\usepackage{subcaption}
\usepackage[font=footnotesize,labelformat=simple]{subcaption}

\newenvironment{itemizer}{\begin{itemize}\setlength{\parsep}{0cm}\setlength{\itemsep}{-0.7em}}{\end{itemize}}
\newenvironment{enumerator}{\begin{enumerate}\setlength{\parsep}{0cm}\setlength{\itemsep}{-0.7em}}{\end{enumerate}}
\newcommand{\OurScheme}{\textsc{SIRA}}
\iffalse
\newcommand{\TODO}[1]{\todo[inline]{#1}}
\newcommand{\hms}[1]{\textcolor{blue}{HMS: #1}}
\newcommand{\danilowi}[1]{\textcolor{magenta}{danilowi: #1}}
\newcommand{\christoph}[1]{\textcolor{orange}{Christoph: #1}}
\newcommand{\felix}[1]{\textcolor{cyan}{Felix: #1}}
\newcommand{\babis}[1]{\textcolor{purple}{Babis: #1}}
\newcommand{\jpk}[1]{\textcolor{teal}{JPK: #1}}
\newcommand{\ian}[1]{\textcolor{olive}{[IC: \textit{#1}]}}

\newcommand{\yaman}[1]{\textbf{YU: #1}}
\newcommand{\tk}[1]{\textcolor{magenta}{TK: #1}}
\else
\newcommand{\TODO}[1]{}
\newcommand{\hms}[1]{}
\newcommand{\danilowi}[1]{}
\newcommand{\christoph}[1]{}
\newcommand{\felix}[1]{}
\newcommand{\babis}[1]{}
\newcommand{\jpk}[1]{}
\newcommand{\ian}[1]{}

\newcommand{\yaman}[1]{}
\newcommand{\tk}[1]{}
\fi

\newacro{fdna}[FDNA]{FPGA dataflow neural network accelerator}
\newacro{dnn}[DNN]{deep neural network}
\newacro{qnn}[QNN]{quantized neural network}
\newacro{mac}[MAC]{multiply-accumulate}
\newacro{ptq}[PTQ]{post-training quantization}
\newacro{qat}[QAT]{quantization-aware training}
\newacro{mpe}[MPE]{matrix of processing engines}
\newacro{llm}[LLM]{large language model}
\newacro{hls}[HLS]{high-level synthesis}
\newacro{pot}[PoT]{power-of-two}

\newcommand{\acrshort}[1]{\acs{#1}}
\newcommand{\acrfull}[1]{\acf{#1}}

\definecolor{siraint}{HTML}{EA6B66}

\newcommand{\varmin}[1]{\underline{#1}}
\newcommand{\varmax}[1]{\overline{#1}}
\newcommand{\ivar}[1]{[\underline{#1},\overline{#1}]}
\newcommand{\siz}[1]{q_{#1}}
\newcommand{\sis}[1]{s_{#1}}
\newcommand{\sib}[1]{b_{#1}}

\newcommand{\sivarnew}[1]{\sis{#1} \cdot [\ii{\siz{#1}},\ia{\siz{#1}}] + \sib{#1}}
\newcommand{\sivarconst}[1]{\sis{#1} \cdot \siz{#1} + \sib{#1}}
\newcommand{\ii}[1]{\underline{#1}}
\newcommand{\ia}[1]{\overline{#1}}
\newcommand{\myvec}[1]{\mathbf{#1}}
\newcommand{\miv}[2]{\mathrm{miv}(\myvec{#1},\myvec{#2})}
\newcommand{\mav}[2]{\mathrm{mav}(\myvec{#1},\myvec{#2})}

\newcommand*\circled[1]{\tikz[baseline=(char.base)]{
            \node[red, shape=circle,draw,inner sep=1pt] (char) {#1};}}

\newcommand{\quantnode}{\texttt{Quant}}

\begin{document}

\title{\OurScheme{}: Scaled-Integer Range Analysis for \\ Optimizing FPGA Dataflow Neural Network Accelerators}

\graphicspath{{figures/}}
\DeclareGraphicsExtensions{.pdf}
\pgfplotsset{compat = 1.3}

\author{Yaman Umuroglu}
\email{yamanu@amd.com}
\orcid{0000-0002-3700-5935}
\affiliation{%
  \institution{AMD Research}
  \city{Trondheim}
  \country{Norway}
}

\author{Christoph Berganski}
\email{christoph.berganski@upb.de}
\orcid{0009-0006-6956-0657}
\author{Felix Jentzsch}
\email{felix.jentzsch@upb.de}
\orcid{0000-0003-4987-5708}
\affiliation{%
  \institution{Paderborn University}
  \city{Paderborn}
  \country{Germany}
}

\author{Michal Danilowicz}
\email{danilowi@agh.edu.pl}
\orcid{0000-0001-8851-8186}
\author{Tomasz Kryjak}
\email{tomasz.kryjak@agh.edu.pl}
\orcid{0000-0001-6798-4444}
\affiliation{%
  \institution{AGH University of Krakow}
  \city{Krakow}
  \country{Poland}
}

\author{Charalampos Bezaitis}
\email{charalampos.bezaitis@ntnu.no}
\orcid{0000-0002-7905-8357}
\author{Magnus Sjalander}
\email{magnus.sjalander@ntnu.no}
\orcid{0000-0003-4232-6976}
\affiliation{%
  \institution{Norwegian University of Science and Technology}
  \city{Trondheim}
  \country{Norway}
}

\author{Ian Colbert}
\email{icolbert@amd.com}
\orcid{0000-0002-1669-5519}
\affiliation{%
  \institution{AMD}
  \city{San Jose}
  \country{USA}
}

\author{Thomas Preusser}
\email{thomas.preusser@amd.com}
\orcid{0000-0003-3998-7896}
\affiliation{%
  \institution{AMD Research}
  \city{Dreseden}
  \country{Germany}
}

\author{Jakoba Petri-Koenig}
\email{jakobap@amd.com}
\orcid{0009-0002-2457-7819}
\author{Michaela Blott}
\email{mblott@amd.com}
\orcid{0000-0002-7833-4057}
\affiliation{%
  \institution{AMD Research}
  \city{Dublin}
  \country{Ireland}
}

\renewcommand{\shortauthors}{Umuroglu et al.}
\renewcommand{\shorttitle}{SIRA: Scaled-Integer Range Analysis for Optimizing FPGA Dataflow Neural Network Accelerators}

\begin{abstract}
While neural network quantization effectively reduces the cost of matrix multiplications, aggressive quantization can expose non-matrix-multiply operations as significant performance and resource bottlenecks on embedded systems. 
Addressing such bottlenecks requires a comprehensive approach to tailoring the precision across operations in the inference computation.
To this end, we introduce scaled-integer range analysis (\OurScheme{}), a static analysis technique employing interval arithmetic to determine the range, scale, and bias for tensors in quantized neural networks. 
We show how this information can be exploited to reduce the resource footprint of FPGA dataflow neural network accelerators via tailored bitwidth adaptation for accumulators and downstream operations, aggregation of scales and biases, and conversion of consecutive elementwise operations to thresholding operations.
We integrate \OurScheme{}-driven optimizations into the open-source FINN framework, then evaluate their effectiveness across a range of quantized neural network workloads and compare implementation alternatives for non-matrix-multiply operations.
We demonstrate an average reduction of 17\% for LUTs, 66\% for DSPs, and 22\% for accumulator bitwidths with \OurScheme{} optimizations, providing detailed benchmark analysis and analytical models to guide the implementation style for non-matrix layers. 
Finally, we open-source \OurScheme{} to facilitate community exploration of its benefits across various applications and hardware platforms.
\end{abstract}

\maketitle

\section{Introduction}

\Acp{dnn} serve critical roles in areas like language understanding, computer vision, and autonomous systems owing to their impressive capabilities to find patterns in large amounts of data with great accuracy.
Their integration into embedded systems enables intelligent data processing and decision-making close to the data source, reducing latency and bandwidth usage while supporting real-time applications with enhanced privacy and autonomy in various scenarios.
However, inference of DNNs requires significant memory and compute capabilities, which is at odds with the stringent system-level requirements of edge computing demanding efficient use of limited resources, low power consumption, and quick processing times.
This has placed increased emphasis on techniques to reduce \textit{inference cost}, which may be measured in time, energy, or other resources.
To reduce inference cost, a wide array of techniques from both the workload and hardware perspectives has emerged, focusing primarily on \ac{mac} operations.
On the workload side, quantization of model weights and activations~\cite{krishnamoorthi2018quantizing, frantar2023optq, zhang2024magr} is highly popular, leading to \acp{qnn} with reduced memory and computational requirements.
From the hardware perspective, custom architectures have emerged that focus on exploiting the plentiful parallelism available in DNNs with dedicated matrix and vector datapaths, as well as enabling practical gains from workload optimizations by supporting reduced-precision datatypes.
The rapidly changing nature of the algorithms and datatypes demands a great deal of hardware flexibility in addition, making FPGAs an attractive option for \ac{qnn} deployment in embedded systems.
In particular, \acp{fdna} that tailor precision and parallelism at the granularity of each layer of a \ac{qnn} have been shown~\cite{duarte2018fast, blott2018finn, nn2fpga} to offer highly efficient solutions.

\begin{table}[h!]
    \centering
    \caption{CIFAR-100 ResNet-8 \acs{qat} top-1 accuracy\% with different bitwidths and scaling factors. \texttt{float32} accuracy is 70.52\%.}
    \begin{tabular}{ccccc}
        & \multicolumn{1}{c}{Power-of-Two Scale} & \multicolumn{2}{c}{Floating-Point Scale} \\ 
         \cmidrule(lr){2-2}         \cmidrule(lr){3-4}
         Quantization & Weight Per Tensor & Weight Per Tensor & Weight Per Channel \\ 
         \midrule
         4-bit & 69.78 & 70.55 & 70.45 \\ 
         3-bit & 67.70 & 69.15 & 70.13 \\ 
         \bottomrule
     \end{tabular}
     \label{tbl:scaling_richness}
\end{table}

While the focus on reducing the cost of \acp{mac} through quantization has brought significant benefits for \ac{qnn} inference, Amdahl's Law casts a shadow over further inference cost reduction by weight and activation quantization.
Several recent works have noted that, as the matrix multiplication precision is reduced, the bottleneck in inference cost starts shifting to non-MAC operations.
Neseem et al.~\cite{neseem2024pikelpn} estimate that elementwise operations such as quantization scaling, activation functions, and batch normalization account for nearly 89\% of the total energy used by a MobileNet-V2 model quantized to 1-bit weights and activations. 
Similarly, Karami et al.~\cite{karami2024nongemm} observe that, as matrix multiplications are accelerated via quantization, elementwise operations contribute significantly to the end-to-end latency, increasing from 17\% to 42\% on average.
This shift of the inference cost bottleneck to non-matrix operations is exacerbated by the fact that more expressive scale factors may be needed as weights and activations are quantized to fewer bits.
A motivating example is provided in \autoref{tbl:scaling_richness},
where a ResNet-8 topology is trained to use 4- and 3-bit weights and activations, with different constraints on the scale factors.
We observe that although 4-bit quantization is reasonably robust, %
the 3-bit case exhibits a difference of over two percentage points between the least
and most
expressive scale factors.
To preserve high accuracy with 3-bit quantization, per-channel floating-point scale factors are required. %
The combined cost of these non-MAC operations, which we refer to as the \emph{layer tail} (\autoref{sec:sira-typical-qnn-example}), can be significant.
To holistically optimize inference, a well-founded reduction of layer tail costs without adversely affecting accuracy is essential.

Towards this end, the ability to determine the value range for each intermediate tensor of a \ac{qnn} is crucial.
By understanding the possible minimum and maximum values, an \ac{fdna} can tailor the bitwidth for storage and computation, enabling resource reductions. 
Furthermore, more sophisticated optimization opportunities that enable balancing memory and computation trade-offs can be identified, such as fusing a sequence of non-MAC operations into a single thresholding operator, unlocking one of the key strengths of FPGAs.
In this paper, we propose scaled-integer range analysis (\OurScheme{}) to perform static analysis of trained \acp{qnn} to extract this valuable information. 
\OurScheme{} employs interval arithmetic to track the possible value ranges of tensors as they propagate through the network's layers. 
By providing integer range, scale, and bias information for each tensor at \ac{fdna} generation time, \OurScheme{} lays the foundation for a more holistic optimization strategy for \ac{qnn} inference on \acp{fdna}, allowing for tailored hardware implementations that efficiently handle both \ac{mac} and non-\ac{mac} operations.
Our main contributions can be summarized as follows:
\begin{itemizer}
    \item We introduce \OurScheme{}, which applies interval arithmetic to \acp{qnn} to enable inference optimization opportunities for \acp{fdna} by providing integer range, scale, and bias information for each tensor.
    We also identify the conditions under which scaled-integer tensors can propagate in a \ac{qnn} to help guide quantization co-design decisions.
    \item We implement a significantly improved version of the \textit{streamlining} optimization~\cite{umuroglu_jahre:CASES2017} based on \OurScheme{}, aggregating scales and biases in layer tails, and optionally converting groups of elementwise operations to thresholds. 
    We offer a set of microbenchmark results and analytical resource cost models to help pick the most effective implementation style for layer tails.
    \item We show how accumulator widths can be minimized with \OurScheme{} for matrix multiplication and convolutional layers for \acrshort{fdna}s, leading to 22\% smaller accumulators on average and propagating resource savings into following non-matrix layers.
    \item Integrating \OurScheme{} and new hardware kernels to take advantage of the optimizations into the FINN \ac{fdna} compiler, we evaluate the effects of our optimizations on a set of \ac{qnn} workloads, demonstrating an average reduction of 17\% in LUTs and 66\% in DSPs.
    \item To help the broader \ac{fdna} ecosystem adopt these techniques, we provide an open-source implementation of \OurScheme{} at \url{https://bit.ly/sira-qonnx}.
\end{itemizer}

\christoph{The rest of the paper is structured as follows... usually this is at the end of the introduction (we could opt for section references in the bullet points above, though?)}

The rest of the paper is structured as follows. 
\autoref{sec:background} introduces relevant background for the paper.
\autoref{sec:sira} details our range analysis technique, and \autoref{sec:fdna-optimizations} describes the \ac{fdna} compiler optimizations we implement on top of \OurScheme{}.
\autoref{sec:finn-integration} presents how \OurScheme{} and optimizations are integrated into the FINN \ac{fdna} compiler.
\autoref{sec:methodology} describes our evaluation methodology for \ac{qnn} workloads and microbenchmarks, followed by \autoref{sec:results} which presents the results of our evaluation.
\autoref{sec:related-work} presents the related work, and \autoref{sec:conclusion} concludes the paper, presenting future directions.

\section{Background}
\label{sec:background}

We introduce relevant background on neural network quantization, FPGA \ac{dnn} accelerators, and interval arithmetic.

\subsection{Quantizing Neural Networks}
\label{sec:qnn-background}

Quantization is the process of representing real-valued tensors with low-precision data formats. When used to reduce neural network inference costs, quantization constraints are commonly separated into two paradigms: (1) weight-only quantization, where only weights are quantized to reduce the cost of storing and/or transferring data; and (2) weight-activation quantization, where both operands of the matrix multiplication are quantized to also reduce the cost of computation.
Weight-activation quantization is predominantly used in resource-constrained inference settings such as embedded or edge computing, where the cost of high-precision computations would be prohibitive.

Quantized models are broadly obtained via either post-training quantization~(PTQ) or quantization-aware training~(QAT). PTQ methods calibrate pre-trained models with little to no data with the goal of minimizing functional deviation from a reference model. QAT methods train or fine-tune models while emulating quantization in the forward pass with the goal of recovering accuracy lost to quantization error via gradient descent, as such QAT methods often yield higher quality quantized models than PTQ methods. 
The output of both processes, however, is a compute graph with quantization operators that emulate low-precision arithmetic via scaling, shifting, rounding, and clipping.

The standard quantization operators, referred to as quantizers, build on uniform affine transformations that map high-precision values (e.g., \texttt{float32}) to low-precision data formats (e.g., \texttt{int4}) via the composition $\mathcal{Q}(x) := ( f \circ g \circ f^{-1} )(x)$, where $f(x) := s \cdot \left( x - z \right)$ and $g(x) := \text{clip}\left( \left\lceil x \right\rfloor; q_{\min}, q_{\max} \right)$.
Here, the function $f$ (sometimes referred to as \textit{dequantization}) and its inverse $f^{-1}$ bridge the domains of integers and real numbers via the \textit{zero point} $z$ and \textit{scaling factor} $s$ while
$g$ maps the scaled and shifted real numbers to integers of
target bitwidth $b$, which is controlled by clipping the scaled and shifted data range to the closed interval $[q_{\min}, q_{\max}]$, where $q_{\max} - q_{\min} = 2^b - 1$ for $b$ bits. %
When quantizing to signed integers, it is common that $[q_{\min}, q_{\max}] = [-2^{b -1}, 2^{b - 1} - 1]$ and $z = 0$. %
When quantizing to unsigned integers, it is common that $[q_{\min}, q_{\max}] = [0, 2^b - 1] $ and $z = \min(x)$. %
Intuitively, $z \neq 0$ is more flexible and enables better alignment to low-precision data format with skewed distributions, but it is also more expensive~\cite{zhang2022learning}.
Therefore, it is common that $z = 0$ for weights and $z \neq 0$ for activations to improve flexibility with reduced overheads~\cite{zhang2022learning}.

The scaling factor $s$ is often defined as the ratio given by
$s = {\max(\vert \bm{x} \vert )}/ {( 2^{b-1}-1 )}$,
where $\max(\vert\bm{x}\vert)$ is the maximum elementwise magnitude in the tensor $\bm{x}$, referred to as per-tensor scaling.
It is increasingly common to use unique scaling factors for each output channel of a layer~\cite{nagel2019data}, or even for unique subgroups of elements within an output channel of a layer~\cite{frantar2023optq}, respectively referred to as per-channel or per-group scaling.
The granularity of the scaling factors presents a critical trade-off: finer granularity tends to preserve model quality better~\cite{frantar2023optq}, but also tends to increase compute and memory costs.
Indeed, Neseem et al.~\cite{neseem2024pikelpn} estimate that per-group scaling increases compute costs by 16-30\% when compared to per-tensor scaling on a 1-bit quantized ResNet-50 model with \texttt{float32} scaling factors.
However, if scaling factors are instead constrained to the form $2^{-i}$, where $i$ is an integer, then one can re-scale using simple bit-shifts without the overhead of floating-point math.
This is referred to as \ac{pot} scaling~\cite{neseem2024pikelpn}.

One can observe across recent studies that finer-grained, higher-precision scaling factors help maintain model quality when quantizing to lower-precision data formats.
Krishnamoorthi~\cite{krishnamoorthi2018quantizing} was one of the first to show this on MobileNet-v2 by restoring nearly +70\% top-1 accuracy on ImageNet by applying per-channel scaling rather than per-tensor scaling while quantizing to 8-bit weights and activations. \christoph{We should reference the accuracy for this as well?} \ian{I think a reference is enough. There is a whole data table on this in the referenced paper, and we are actively trying to trim the background section.}
More recently, Frantar et al.~\cite{frantar2023optq} show that when quantizing the weights of OPT-175B to 2 bits with per-group scaling, decreasing the group size (i.e., increasing the scaling granularity) from 1024 to 32 elements improves model perplexity almost as much as increasing the bitwidth to 3 bits.
Zhang et al.~\cite{zhang2024magr} show this is especially pronounced in ``smaller'' models, where scaling 2-bit quantized weights in groups of 128 elements improves Llama2 7B perplexity 60\% more than that of Llama2 70B when compared to per-channel scaling.
We support these results with our motivating experiments in \autoref{tbl:scaling_richness}, where we highlight that power-of-two quantization results in higher classification error as the operand width decreases.

\subsection{FPGA Neural Network Accelerators}
\begin{figure}[htbp]
    \centering
    \begin{subfigure}{0.5\textwidth}
        \centering
        \includegraphics[height=1.2in]{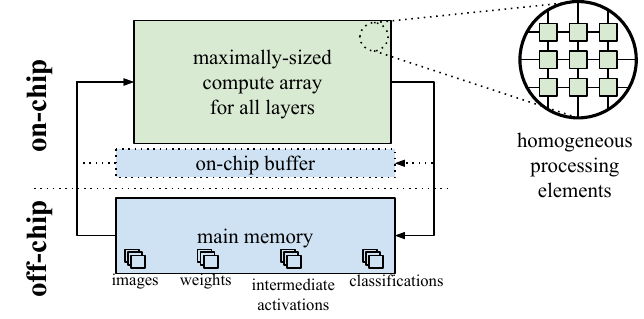}
        \caption{Matrix of Processing Engines (MPE)-Style Accelerator}
        \label{fig:mpe}
    \end{subfigure}
    ~ 
    \begin{subfigure}{0.5\textwidth}
        \centering
        \includegraphics[height=1.2in]{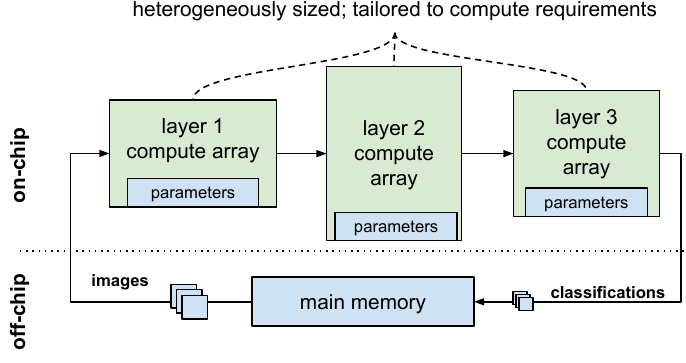}
        \caption{FPGA Dataflow Neural Network Accelerator (\acrshort{fdna})}
        \label{fig:df}
    \end{subfigure}
    \caption{Architecture styles for FPGA-based neural network inference acceleration, adapted from~\citet{umuroglu2017finn}}
    \label{fig:arch_types}
\end{figure}
Two concepts have emerged to leverage the parallelism of the FPGA fabric for neural network inference as illustrated in \autoref{fig:arch_types}:
\autoref{fig:mpe} the more generic architecture with a matrix of processing engines (MPE), and \autoref{fig:df} the highly customized dataflow architecture, referred to as \acrfull{fdna}.

MPE-style accelerators offer an architecture tailored to typical neural network operations, such as an array of multiply-and-accumulate (MAC) units, providing a fixed degree of parallelism over the maximally-sized compute array.
This architecture is %
designed to run any DNN, which gives it a high degree of flexibility compared to the \acrshort{fdna} approach.
However, this flexibility comes at a cost; the layer-by-layer execution requires buffering of the intermediate results. To achieve high throughput, the data is typically processed in batches. Buffering limits the minimal latency at which an individual data sample can be processed. Furthermore, the fixed architecture may be underutilized since it is typically designed to support the most compute-intensive layer. %

When implementing \acrshort{fdna}s, each layer is mapped to the FPGA and occupies its own compute and memory resources, forming a deep processing pipeline.
Data is continuously streamed in instead of doing batch processing.
Resources are allocated according to the requirements of each individual layer; %
this architecture style allows for a custom-tailored, highly specialized and optimized accelerator design taking advantage of the configurability of the FPGA fabric.
DNN inference in this style can achieve very low latency and high throughput, but is also more limited by the available resources. The high degree of customization means that the architecture needs to be redesigned for each neural network.
Both architecture styles have been explored in prior work to implement neural network inference on FPGAs, and compiler stacks have emerged that facilitate the development process, such as hls4ml~\cite{duarte2018fast}, NN2FPGA~\cite{nn2fpga}, and FINN~\cite{umuroglu2017finn, blott2018finn}. 

\subsection{QONNX}
\label{sec:qonnx}
\begin{figure}[htbp]
  \centering
  \includegraphics[width=0.75\linewidth]{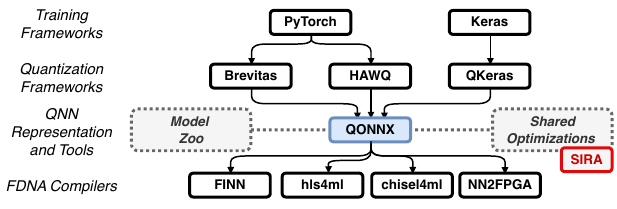}
  \caption{The QONNX ecosystem, enabling \acrshort{fdna} compilers to share QNN representation and optimizations, such as SIRA.} %
  \label{fig:qonnx_ecosystem}
\end{figure}
QONNX \cite{qonnx_paper, qonnx_repo} builds on top of the standard ONNX representation, extending it by custom operators such as the \quantnode{} operator illustrated in \autoref{fig:sira_quant_example}, which represents arbitrary bitwidth integer quantization.
\quantnode{} takes in four inputs: the input to be quantized, the scale factor $s$, the zero-point $z$, and the bitwidth $b$.
Two additional attributes are available to adjust the quantization range $[q_{\min}, q_{\max}]$: \textit{signed}, indicating whether the output is signed or unsigned integers, and \textit{narrow}, which is only valid when \textit{signed} is enabled and sets $q_{\min} = -2^{b-1}-1$.
This enables taking advantage of the fine-grained flexibility of the FPGA fabric, and several \acrshort{fdna} compilers, including FINN~\cite{umuroglu2017finn,blott2018finn}, hls4ml~\cite{duarte2018fast}, chisel4ml~\cite{chisel4ml}, and NN2FPGA~\cite{nn2fpga} support QONNX as their input or intermediate representation.
QONNX is exported from several quantization frameworks, such as Brevitas~\cite{brevitas} and HAWQ~\cite{campos2023end}, which are based on PyTorch, and QKeras~\cite{coelho2021automatic}, which is based on Keras. This ecosystem of training and quantization frameworks, as well as \acrshort{fdna} compilers, is shown in \autoref{fig:qonnx_ecosystem}. Besides this ecosystem integration, QONNX provides a model zoo of pretrained example neural networks alongside a library of shared, non-backend-specific optimizations, such as constant folding and data layout conversion. \OurScheme{} is proposed as an addition to this set of shared optimizations.

\subsection{Interval Arithmetic}
\label{sec:intervaL_arithmetic}

Interval arithmetic is a mathematical technique that, given input bounds, enables the computation of \textit{guaranteed} output bounds for a particular function~\cite{moore2009introduction}.
In interval arithmetic, a variable $v$ is expressed as an interval $\ivar{v}$, where $\ivar{v}$ denotes the interval with minimum $\underline{v}$ and maximum $\overline{v}$ values for the variable $v$. While interval arithmetic offers guaranteed bounds, the bounds may be loose due to the dependency problem~\cite{dawood2019logical}.
Previously, interval arithmetic has been used in \ac{hls} approaches (\autoref{sec:related-work-rangeanalysis}) for optimization, and for verification of adversarial robustness in neural networks in terms of the sensitivity of the neural network to small changes in the input~\cite{gowal2018effectiveness}.
Given a function $y = f(v)$, the interval for the output $\ivar{y}$ is computed by combining the input intervals in specific ways, depending on the function $f$.
We briefly cover this for two key classes of neural network functions below.

\subsubsection{Element-Wise Monotonic Functions}
\label{sec:element_monotonic}

If the element-wise local minimum and maximum of the function output are generated by extrema (minimum or maximum) of the input space, we consider it \textit{element-wise monotonic}. These corners correspond to combinations of the minimum and maximum values of the input variables. This property is satisfied by many types of neural network layers, including ReLU and sigmoid activation functions, concatenation, batch normalization, element-wise addition and multiplication, max, and average pooling, as well as quantization operators. 
\subsubsection{Constant-Weighted Dot-Product Functions}
\label{sec:cwt_dotprod}

For functions that involve constant-weighted dot products (e.g., fully-connected or convolutional layers with fixed weights) the element-wise monotonic property (\autoref{sec:element_monotonic}) does not apply.
Here, a simplified version of affine-layer propagation as presented by Gowal et al.~\cite{gowal2018effectiveness} can be applied, using \textit{minimizing} and \textit{maximizing} input vectors, respectively denoted as $\miv{w}{i}$ and $\mav{w}{i}$.
To construct the minimizing input vector, the maximum of the input range for each negative weight and the minimum for each positive weight are used. %
The maximizing vector is constructed vice versa.
To obtain the range of the outputs, the dot product of the weights with the minimizing and maximizing input vectors is computed.

\section{\OurScheme{}: Scaled-Integer Range Analysis for Neural Networks}
\label{sec:sira}

We now introduce \OurScheme{}, which performs \textit{scaled-integer range analysis} on trained QNNs.
In neural networks with integer-quantized weights and activations, most (but not all) tensors have an underlying integer component with a fixed affine relationship to the full-precision tensor.
We can express their ranges as \emph{scaled-integer range} (i.e., an integer range with a scale and bias).
Specifically, $\ivar{v} = \sivarnew{v}$, where $\ivar{\siz{v}}$ is the integer range with scale $\sis{v}$ and bias $\sib{v}$. 
Both $\sis{v}$ and $\sib{v}$ must have shapes that can be broadcast to the tensor shape, and may (or may not) have a restriction such as being a power-of-two. 
With \OurScheme{}, we require $\sis{v}$ and $\sib{v}$ to be constants, since allowing scales and biases to be ranges themselves significantly increases the complexity of the required analysis and results in very wide ranges, from which it is difficult to extract useful insights.
As part of our analysis, we are interested in obtaining the following information for a given tensor from a trained QNN:
\begin{enumerator}
    \item What is the possible value range $\ivar{v}$ for the tensor $v$?
    \item If there is an underlying integer component for the tensor $v$, what is its range $\ivar{\siz{v}}$ and the affine transform with scale and bias $\ivar{v} = \sivarnew{v}$ to obtain it?
    \item If there is an underlying integer component for the tensor $v$, which other tensors have contributed to the scale and bias terms?
\end{enumerator}

\subsection{Overview}
\label{sec:sira_overview}

Our end goal is to apply \OurScheme{} to real-world QNNs with a view towards optimizing \acrshort{fdna} deployment. To achieve this goal, we need to be able to accurately represent different operators, examine range information for their inputs, and propagate range, scale, and bias information accordingly. 
Towards this end, we implement \OurScheme{} in QONNX as a shared optimization (\autoref{sec:qonnx}).
Throughout the paper, we provide a combination of mathematical formulae and illustrative examples that uses operator definitions from the (Q)ONNX standard~\cite{qonnx_paper} to help describe how \OurScheme{} works.

\begin{minipage}{\linewidth}
\begin{lstlisting}[basicstyle=\footnotesize,language=Python,label=lst:node-by-node,numbers=left,escapechar=|,numbersep=-6pt,caption={Node-by-node scaled-integer range propagation.}]
    class ScaledIntRange:
        range: tuple(array, array) # full precision min, max range
        int_range: None or tuple(array, array) # integer min, max range
        scale: None or array # scale to go from int_range to range
        bias: None or array # bias to go from int_range to range
        
    def SIRA(graph, input_names[], input_ranges[]: List[ScaledIntRange]):
        range{} = {input_names[] : input_ranges[]}|\label{lst:node-by-node:range-dict}| 
        for node in graph:
            input_ranges[] = range[node.input[]]|\label{lst:node-by-node:input-range}|
            ouput_ranges[] = node.propagate(input_ranges[])|\label{lst:node-by-node:ouput-range}|
            range[node.output[]] = output_ranges[]|\label{lst:node-by-node:dict-update}|
        return range
\end{lstlisting}
\end{minipage}

\OurScheme{} is built on the foundation of applying interval arithmetic (\autoref{sec:intervaL_arithmetic}) to the neural network graph, enhanced with optional scale and bias information to distinguish any underlying integer component.
As shown in \autoref{lst:node-by-node}, this is achieved by a node-by-node walk of the topologically sorted QONNX graph representing the QNN.
Our range analysis takes the QONNX \texttt{graph}, a list of named input values (\texttt{input\_names[]}), and a list of corresponding value ranges (\texttt{input\_ranges[]}) as input.
A \texttt{range} dictionary is initially populated with the value ranges of the input values (line~\ref{lst:node-by-node:range-dict}).
Then, for each node, the input range information is fetched from the dictionary (line~\ref{lst:node-by-node:input-range}), a node-dependent propagation function is called to compute the output range information for this node (line~\ref{lst:node-by-node:ouput-range}), and the dictionary is updated with the newly calculated value range (line~\ref{lst:node-by-node:dict-update}).
Different node types require different propagation handlers, and we present the different handlers below.

\subsection{Propagation for Scaled-Integer Ranges}
\label{sec:sira-propaget}

We can only propagate scaled-integers from inputs of an operation to its outputs in certain cases, depending on a combination of the operation type and the properties of input ranges.
We start by identifying the general rules by which scaled-integer propagation works.

\begin{itemizer}
    \item If an operation does not propagate scales and biases, they are left unspecified, and the output tensor is assumed not to be scaled-integer. In this case, we only propagate ranges as described by prior art; see \autoref{sec:intervaL_arithmetic}.
    \item Non-linear operations do not propagate scales and biases, unless covered by an exception below.
    \item Scales and biases only propagate in affine regions of the graph (i.e., regions with only multiplication and addition).
    \item For scale and bias propagation to be possible, at least one of the dynamic inputs to the operation must be scaled-integer. The quantization operation is an exception and always outputs scaled-integer ranges.
    \item Additional constraints on the granularity or values of scales and biases may be required, as detailed in the following subsections.
    \item If an operation not specified here can be \textit{lowered} to one or more of the operations below, it can be handled in this manner. For instance, \texttt{BatchNormalization} can be lowered to a multiplication and an addition. 
\end{itemizer}

\subsubsection{Quantization}
\label{sec:sira-quantization}
Quantization operators \emph{enforce} the scaled-integer ranges defined by their inputs, as per the definition in \autoref{sec:qonnx}.
\ian{Reading this again, I think that the definition can be made clearer. My concern is that the general quantizer doesn't output a range, but we are equating them. Hope my attempt below helps.}
Let $\sivarnew{\myvec{x}}$ be the scaled-integer range resulting from quantizing tensor $\myvec{x}$ to $q_{\myvec{x}}$ via $\mathcal{Q}(\myvec{x}) = \myvec{s} \cdot ((g \circ f^{-1})(\myvec{x}) - \myvec{z})$, with quantization scale $\myvec{s}$ and zero-point $\myvec{z}$. 
The components of the scaled-integer range can be computed as $\sis{x} = \myvec{s}, \sib{x} = - \myvec{s}\cdot\myvec{z}, \varmin{\siz{x}} = (g \circ f^{-1})(\varmin{\myvec{x}}), \varmax{\siz{x}} = (g \circ f^{-1})(\varmax{\myvec{x}})$.

\begin{figure}
  \centering
  \includegraphics[height=4cm]{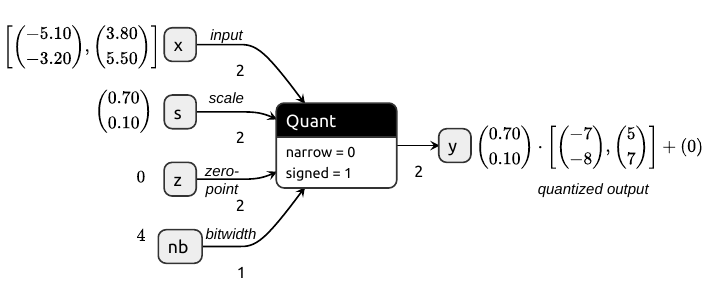}
  \caption{\OurScheme{} example for a \texttt{Quant} node. signed = 1 and narrow = 0 indicate signed quantization and no narrow-range (\autoref{sec:qonnx}). 
  Numbers on arrows indicate tensor shapes, and numbers on node inputs and outputs indicate inputs to and outputs from \OurScheme{}.}
  \label{fig:sira_quant_example}
\end{figure}

This is exemplified in \autoref{fig:sira_quant_example}, with per-channel, non-integer input range and per-channel scales. Note how the output is a scaled-integer range with the scale factor set to the \texttt{Quant} node scale factor, and no bias in this case since the zero-point input is set to zero. 
In this example, %
the output integer range of the first channel is $[-7, 5]$, which does not span the full range of 4 bits, which is  $[-8, 7]$. 

\subsubsection{Addition}

We identify two cases under which we can propagate scaled-integer ranges through addition. Let $v = \sivarnew{v} = v_0 + v_1$.
\begin{enumerate}
    \item If input $v_0=\sivarnew{v_0}$ is a scaled-integer and $v_1$ is a constant, we can simply absorb the $v_1$ into the output bias regardless of whether it is a scaled-integer or not. In this case, $\sis{v}=\sis{v_0}, \sib{v}=\sib{v_0}+v_1, \siz{v}=\siz{v_0}$.
    \item If both inputs are scaled-integers, and their scale factors have an integer relationship (\textit{i.e.}, $\sis{v_1} = k \cdot \sis{v_0}, k \in \mathbb{Z}$, then $\ivar{\siz{v}} = [\ii{\siz{v_0}} + k \cdot \ii{\siz{v_1}}, \ia{\siz{v_0}} + k \cdot \ia{\siz{v_1}} ]$ and $\sis{v} = \sis{v_0}, \sib{v} = \sib{v_0} + \sib{v_1}$).
\end{enumerate}

\begin{figure}
  \centering
    \begin{subfigure}[b]{0.95\textwidth}
        \centering
        \includegraphics{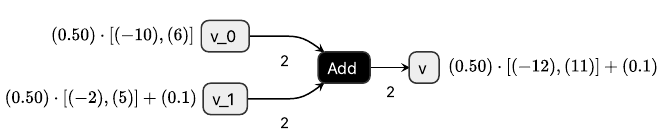}
        \caption{\texttt{Add} node, with matching scales on inputs.}
        \label{fig:sira_add_example}
    \end{subfigure}
    \begin{subfigure}[b]{0.95\textwidth}
        \centering
        \includegraphics{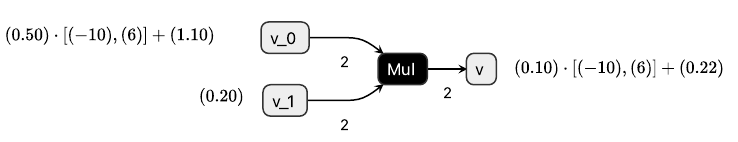}
        \caption{\texttt{Mul} node with a constant on one input.}
        \label{fig:sira_mul_example}
    \end{subfigure}
  \caption{\OurScheme{} example for \texttt{Add} and \texttt{Mul}, indicating the constraints that make scaled-integer propagation possible.}
  
\end{figure}

The example in \autoref{fig:sira_add_example} illustrates the case when $k=1$ (i.e., both inputs have scaled-integer ranges with equal scale) in which case the output integer range becomes a sum of the input integer ranges. 

\subsubsection{Multiplication}

Let $v = \sivarnew{v} = v_0 \cdot v_1$.
To ensure constant $\sis{v}$ and $\sib{v}$ for the output, we require that one input is a scaled-integer $\sivarnew{v_0}$ and the other $v_1$ is a constant. In this case, we can simply apply this constant multiplication to the input scale and bias to obtain the output scale and bias as $\ii{\siz{v}} = \ii{\siz{v_0}}, \ia{\siz{v}} = \ia{\siz{v_0}}, \sis{v} = \sis{v_0} \cdot v_1, \sib{v} = \sib{v_0} \cdot v_1$. The constant input $v_1$ is not required to be an integer, as exemplified in \autoref{fig:sira_mul_example}.

\subsubsection{Matrix Multiplication and Convolution}
\label{sec:sira-matmul-and-conv}
\begin{figure}
  \centering
  \includegraphics{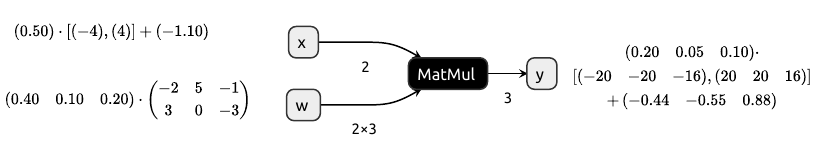}
  \caption{Scaled-integer range analysis example for a \texttt{MatMul} node.}
  \label{fig:sira_matmul_example}
\end{figure}

Consider the multiplication between scaled-integer matrices $Y=W \cdot X$, where $W = \sivarconst{W}$ is the fixed weight matrix of size $M \times K$, and $X = \sivarnew{X}$ is the dynamic input matrix of size $K \times N$. 
To keep a constant scale for each dot product, we must avoid mixing elements with different scales within a dot product, although individual dot products can have different scales.
This implies that the inputs must have at most per-channel scaling (i.e., of shape $M \times 1$ for $\sis{W}$ and $1 \times N$ for $\sis{X}$). %
Fortunately, scaling is commonly applied per-channel for weights and per-tensor for activations~\cite{zhang2022learning}, as per-channel for both sides requires a large $M \times N$ elementwise matrix multiplication for the combined scale.
Additionally, to avoid a non-constant bias for the output, the weights $W$ must have zero bias, although the inputs $X$ can have arbitrary bias.  
When these constraints are satisfied, the output scale is the product of the weight and input scales $\sis{Y}=\sis{W}\cdot\sis{X}$, and the output bias can be computed by multiplying the weight matrix with the input bias broadcast to the input shape $\sib{Y}=W \cdot \sib{X}$.
Finally, the output integer range $\siz{y}$ can be computed as described in \autoref{sec:cwt_dotprod}. 
\autoref{fig:sira_matmul_example} provides an example for propagating a scaled-integer range with both scale and bias through a matrix multiplication with scaled-integer, per-output-channel scaled weights.

We can treat convolutions in the same way, since they can be lowered to matrix-matrix multiplications~\cite{chellapilla2006high}.
An important special case to highlight is depthwise convolution.
Unlike dense convolution, which combines different channels from the input into the same dot product, depthwise convolution has a sparse structure that does not mix different input channels in the same dot product.
In this case, no $M \times N$ elementwise matrix multiplication is needed for the output scale, as a $M \times 1$ scale will suffice.

\subsection{Example: A Typical QNN Layer and Layer Tail}
\label{sec:sira-typical-qnn-example}
\begin{figure}
  \centering
  \includegraphics[width=0.95\linewidth]{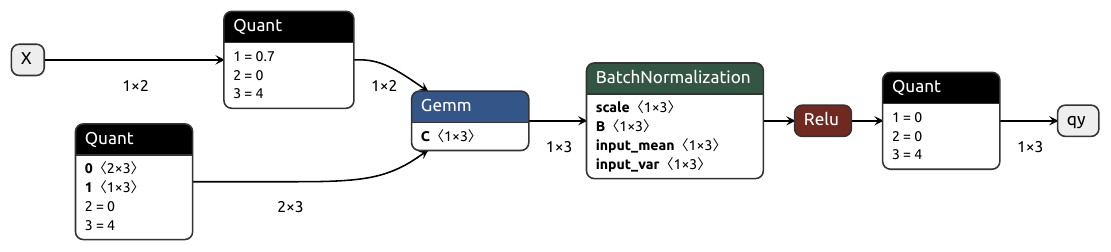}
  \caption{Typical QNN layer prior to lowering of \texttt{Gemm} and \texttt{BatchNormalization}.}
  \label{fig:sira_example_prelower}
\end{figure}

\begin{figure}
  \centering
  \includegraphics[width=0.95\linewidth]{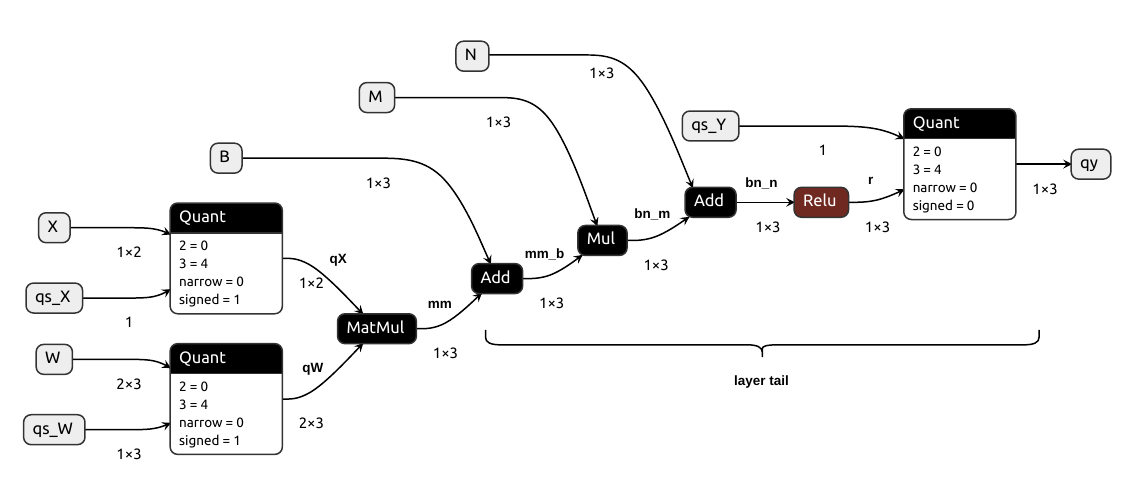}
  \caption{Typical QNN layer after lowering, ready for analysis by \OurScheme{}. \textbf{Bold} indicates intermediate tensor names.}
  \label{fig:sira_example_postlower}
\end{figure}

\autoref{fig:sira_example_prelower} illustrates a typical example of a QNN layer.
Constant-valued tensors are not shown explicitly for brevity. In this example, inputs and weights are quantized to signed 4-bit values with respectively per-tensor and per-channel scales.
They are then fed to a \texttt{Gemm} operator, which also adds a constant bias $C$.
The outputs pass through \texttt{BatchNormalization} and a \texttt{Relu} activation, and finally are quantized to unsigned 4-bit values.
To analyze this graph with \OurScheme{}, we start by lowering particular operations so that they can be processed by the handlers defined above.
Namely, \texttt{Gemm} with a bias is lowered to \texttt{MatMul} and \texttt{Add}, and \texttt{BatchNormalization} is lowered to \texttt{Mul} and \texttt{Add}. 
The resulting graph is shown in \autoref{fig:sira_example_postlower}, which now consists of operations with defined \OurScheme{} handlers and is ready for analysis.
Note the numerous elementwise operations following the \texttt{MatMul}, which constitutes the \textit{layer tail}.

\begin{table}[]
    \centering
    \caption{Example input ranges for \autoref{fig:sira_example_postlower}.}
    \small
    \begin{tabular}{ cc p{4em} cc }
    \cmidrule{1-2}  \cmidrule{4-5}
    Input Tensor & Range  & &
    Input Tensor & Range  \\
    \cmidrule{1-2}  \cmidrule{4-5}
    \addlinespace[-2ex]
    \\
    \texttt{X}  &	$\left[\begin{pmatrix} -5.10 & -3.80 \end{pmatrix}, \begin{pmatrix}  5.10 &  3.80 \end{pmatrix}\right]$ & &
    \texttt{B} & $\begin{pmatrix} -3.30 & -5.20 & -6.10 \end{pmatrix}$ 
    \\
    \texttt{qs\_X} & $0.7$ & &
    \texttt{M} &	$\begin{pmatrix}  0.60 & 0.20 & 0.40 \end{pmatrix}$
    \\
    \texttt{W} &	$\begin{pmatrix} -2.10 & 5.00 & -1.30\\  3.10 & 0.00 & -3.20 \end{pmatrix}$ & &
    \texttt{N} & $\begin{pmatrix} -0.20 & -0.40 & 1.10 \end{pmatrix}$ 
    \\
    \texttt{qs\_W} &	$\begin{pmatrix} 0.20 & 0.30 & 0.10 \end{pmatrix}$ & &
    \texttt{qs\_Y} & $0.10$ 
    \\
    \cmidrule{1-2}  \cmidrule{4-5}
    \end{tabular}
    
    \label{tab:sira_example_input_ranges}
\end{table}

To provide a concrete numerical example, we use the input ranges specified in \autoref{tab:sira_example_input_ranges}.
Note that only $X$ is dynamic; inputs driven by constant values such as trained weights, biases, and so on do not need to be explicitly specified, as they can be automatically inferred as point ranges.
Additionally, in this example, none of the inputs are scaled-integers; these are created by \texttt{Quant} nodes in the graph.
By following the node-by-node propagation described in \autoref{lst:node-by-node} and invoking the appropriate handlers, \OurScheme{} produces the scaled-integer ranges listed in \autoref{tab:sira_example_ranges}.

\begin{table}[]
    \centering
    \caption{Scaled-integer ranges computed by \OurScheme{} for \autoref{fig:sira_example_postlower}. When scale and bias are specified, Range only shows the integer range component $\ivar{\siz{v}}$ as part of the scaled-integer range $\sivarnew{v}$.}
    \small
    \begin{tabular}{ c c c c }
    \toprule
    Tensor & Scale & Range & Bias  \\
    \midrule\\
    \addlinespace[-2ex]
    \texttt{qW} & $\begin{pmatrix} 0.20 & 0.30 & 0.10 \end{pmatrix}$ & $\begin{pmatrix} -8  &  7  & -8 \\ 7  &  0  & -8 \end{pmatrix}$ & 0 \\
    \texttt{qX} & $0.70$ & $\left[\begin{pmatrix} -7  & -5  \end{pmatrix}, \begin{pmatrix} 7  &  5\end{pmatrix}\right]$ & 0 \\
    \texttt{mm} & $\begin{pmatrix} 0.14 & 0.21 & 0.07 \end{pmatrix}$ & $\left[\begin{pmatrix} -91  & -49  & -96  \end{pmatrix}, \begin{pmatrix} 91  &  49  &  96 \end{pmatrix}\right]$ & 0 \\
    \texttt{mm\_b} & $\begin{pmatrix} 0.14 & 0.21 & 0.07 \end{pmatrix}$ & $\left[\begin{pmatrix} -91  & -49  & -96  \end{pmatrix}, \begin{pmatrix} 91  &  49  &  96 \end{pmatrix}\right]$ & $\begin{pmatrix} -3.30 & -5.20 & -6.10 \end{pmatrix}$ \\
    \texttt{bn\_m} & $\begin{pmatrix} 0.08 &  0.04 & 0.03 \end{pmatrix}$ & $\left[\begin{pmatrix} -91  & -49  & -96  \end{pmatrix}, \begin{pmatrix} 91  &  49  &  96 \end{pmatrix}\right]$ & $\begin{pmatrix} -1.98 & -1.04 & -2.44 \end{pmatrix}$ \\
    \texttt{bn\_n} & $\begin{pmatrix} 0.08 & 0.04 & 0.03 \end{pmatrix}$ & $\left[\begin{pmatrix} -91  & -49  & -96  \end{pmatrix}, \begin{pmatrix} 91  &  49  &  96 \end{pmatrix}\right]$ & $\begin{pmatrix} -2.18 & -1.44 & -1.34 \end{pmatrix}$ \\
    \texttt{r} & - & $\left[\begin{pmatrix} 0 & 0 & 0 \end{pmatrix}, \begin{pmatrix} 5.46 &  0.62 &  1.35 \end{pmatrix}\right]$ & - \\
    \texttt{qy} & $0.10$ & $\left[\begin{pmatrix} 0 & 0 & 0 \end{pmatrix}, \begin{pmatrix} 15  &  6  &  13  \end{pmatrix}\right]$ & 0 \\
    \bottomrule
    \end{tabular}
    \label{tab:sira_example_ranges}
\end{table}

\section{Optimizing FPGA Dataflow Neural Network Accelerators With \OurScheme{}}
\label{sec:fdna-optimizations}

The ability of a fine-grained and flexible allocation of FPGA fabric resources enables us to use the range, scale, and bias information obtained by \OurScheme{} to optimize \acrshort{fdna}s in a variety of ways. 
We present two examples for SIRA optimizations. %

\subsection{Streamlining with \OurScheme{}}
\label{sec:streamlining}

\begin{figure}[t]
    \centering
    \includegraphics[width=0.7\textwidth]{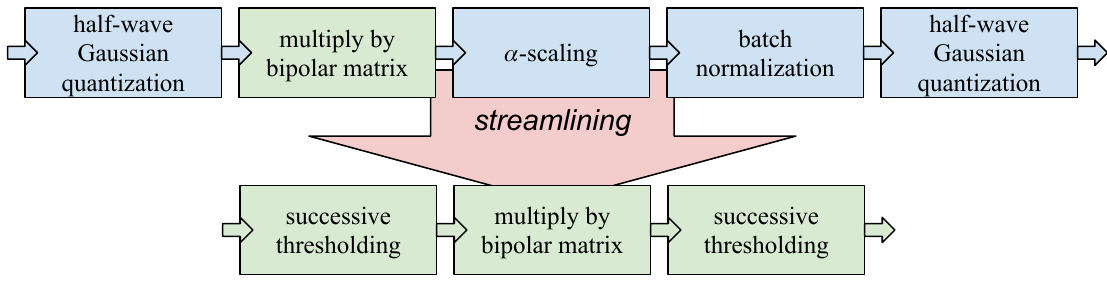}
    \caption{Streamlining example reproduced from \cite{umuroglu_jahre:CASES2017}. Blue and green colors indicate floating-point and integer operations, respectively.}
    \label{fig:original_hwgq_streamlining}
\end{figure}

The non-integer scaling factors and biases applied along the computational graph have the tensors in a QNN not to appear integer at first glance. 
This makes mapping the computations to optimized integer kernels a challenge for \acrshort{fdna} compilers.
In this context, Umuroglu and Jahre~\cite{umuroglu_jahre:CASES2017} describe \emph{streamlining} as a flow to convert all QNN inference operations to integer operations without requiring any additional quantization.
Here, all quantized activations are first converted into multi-thresholding (\autoref{sec:thresholding_conversion}) operations that then \textit{absorb} all non-integer linear transformations by manipulating the threshold values. Thereby, all kernels in the layer tail are fused into a single thresholding layer.
A small example of a QNN inference graph before and after streamlining is shown in \autoref{fig:original_hwgq_streamlining}.
Note how the final graph only consists of integer operations with efficient \acrshort{fdna} mappings.
However, we identify several problems with this streamlining proposal:
\begin{enumerator}
    \item  A concrete streamlining algorithm is missing. It remains unclear how scale factors are identified and moved past other operations. Because of this, existing streamlining implementations (\autoref{sec:handling-scale-factors}) are not generic and need to be customized for individual QNN topologies.
    \item  The conversion of quantized activations into thresholds is only illustrated by a single example.
    \item  The ability to handle scales and biases is completely dependent on the presence of subsequent thresholding that can absorb them. Other scenarios are not supported.
    \item While efficient for few-bit activations, the cost of thresholding still grows exponentially with an increasing activation bitwidth, and linearly with channel count when per-tensor scaling is used. In general, the trade-offs around thresholding are not analyzed.
\end{enumerator}

In the following sections, we resolve these limitations by re-framing the streamlining problem and by discussing how it can be implemented on top of \OurScheme{} as two generic optimizations.

\subsubsection{A Fresh Look at Streamlining.}
We start by noting that the problem of moving scale and bias tensors across different operations, as proposed by \citet{umuroglu_jahre:CASES2017}, is analogous to how \OurScheme{} propagates scales and biases as part of the analysis.
We distinguish this from threshold conversion, which we treat as a separate, optional optimization.
Utilizing the range and scale information, we also formulate a general-purpose threshold conversion procedure that can be adapted for a broad set of quantized activations.
Thus, \OurScheme{}-based streamlining operates in two distinct phases:
\begin{enumerator}
    \item Aggregate multiple scale and bias operators in linear regions into single scale and bias operators, revealing integer \texttt{MatMul} and \texttt{Conv} kernels in the process.
    \item Optionally convert scale, bias, monotonic activation, and activation quantizers into a thresholding layer, collapsing the entire layer tail.
\end{enumerator}

By prioritizing the aggregation of scales and biases and by treating threshold conversion as an optional step, our streamlining implementation is more flexible and able to optimize a broad set of QNN workloads.
We explain all steps in detail in the next sections.

\subsubsection{Aggregating Scales and Biases.} 
\label{sec:aggregate-scale-bias}

The first step in our revised streamlining implementation aggregates scales and biases in linear regions, with two aims: (a) revealing integer kernels for MAC-intensive layers like \texttt{Conv} and \texttt{MatMul} and (b) fusing elementwise additions and multiplications into single \texttt{Mul} and \texttt{Add} operations.
In this process, we choose the position in the graph at which the aggregated scale and bias will be inserted. We call it the \emph{target tensor}, which must follow all MAC operations while remaining within a linear region of the graph.
Note that scales and biases are typically implemented using floating-point data types in neural networks. They do not behave like real numbers due to a lack of associativity and rounding effects. Especially in cases where scales and biases with differences in magnitude are aggregated, numerical errors may be introduced. 
Picking the target tensor later in the linear region will allow more scales and biases to be aggregated, increasing efficiency but also the risk of numerical errors.
In our implementation, we opt for as much aggregation as possible by picking the tensors feeding activation functions as the target tensors, which form the boundary of the linear region.

As \OurScheme{} already separates the integer components from scales and biases for each tensor, the actual aggregation transformation is straightforward and can be performed as follows:
\begin{enumerator}
    \item To enable the manipulation of scales and biases independently, duplicate shared scales and biases that feed multiple consumers. These arise from \texttt{Quant} nodes, where the scale and zero-point operate on both the input and the output, or \texttt{Add} or \texttt{Mul} nodes with outputs branching out to several consumers.
    \item Execute \OurScheme{} while keeping track of which tensors contribute to the scale and bias for each tensor.
    \item Create \texttt{Mul} and \texttt{Add} nodes in front of the target tensor, using respectively the scale and bias values reported for the target tensor by \OurScheme{}.
    \item Erase the original scales and biases, to avoid scaling and biasing twice. Using the contribution history for the target tensor, set all tensors that contribute to the scale and bias to identity values, (i.e., one for scale contributions, and zero for bias contributions).
    \item Remove identity operations such as \texttt{Mul(x,1)} and \texttt{Add(x,0)} to clean up the graph.
\end{enumerator}

\begin{figure}[t]
  \centering
  \includegraphics[width=0.96\textwidth]{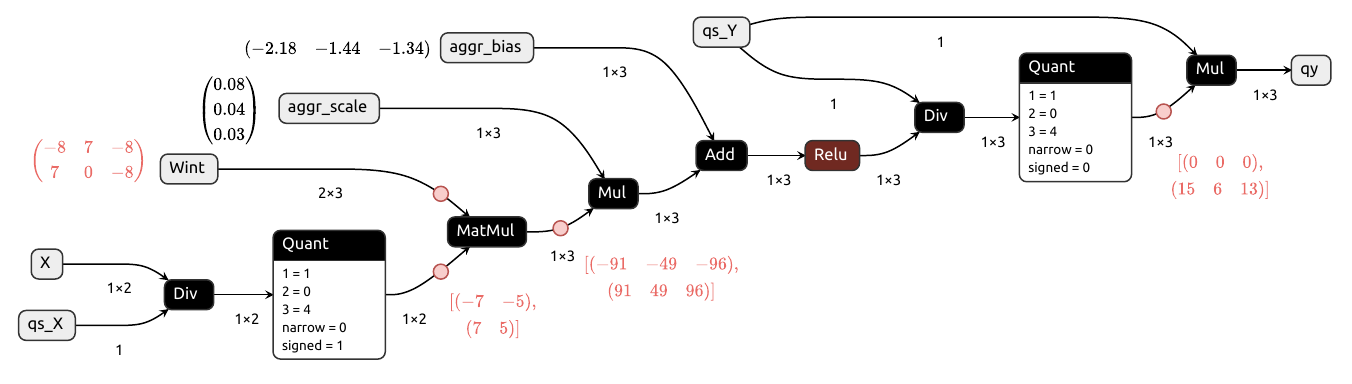}
  \caption{Scales and biases aggregated with \OurScheme{}. {\color{siraint} Red} indicates integer tensors.}
  \label{fig:aggregate_scale_bias}
\end{figure}

Building on top of the previous example in \autoref{sec:sira-typical-qnn-example} for \OurScheme{}, \autoref{fig:aggregate_scale_bias} provides an example of scale and bias aggregation.
This enables a range of optimizations for \acrshort{fdna} compilation, which we identify for this particular example as follows:
\begin{itemizer}
    \item The inputs to and outputs from the \texttt{MatMul} operation are now pure (non-scaled) integers, facilitating the use of an optimized integer compute kernel, and enabling lossless accumulator minimization, as explained in \autoref{sec:accumulator-sizing}.
    \item The various scale and bias factors \texttt{M} and \texttt{N} for \texttt{BatchNormalization}, the scale factor for the weight quantizer \texttt{qs\_W}, and the bias for the \texttt{Gemm} operation \texttt{B} are replaced by a single scale \texttt{aggr\_scale} and bias \texttt{aggr\_bias}. 
    This requires fewer operations executed and parameters stored, compared to the example in \autoref{fig:sira_example_postlower}.
    The aggregated scale and bias can be implemented by floating-point or by fixed-point operations. The latter reduces resource cost at the potential expense of some accuracy.
    \item Scales for the quantizers for input \texttt{qs\_X} and output \texttt{qs\_Y} activations are left intact. Since their scaling operations are made explicit through \texttt{Div} and \texttt{Mul} operations, the scale for the \texttt{Quant} nodes themselves is set to 1, and pure (non-scaled) integer outputs are visible. Although not shown here, the final \texttt{Mul} node can be absorbed into the following layer's scale computation and subsequently removed.
\end{itemizer}

\subsubsection{Converting Quantized Functions to Thresholds}
\label{sec:thresholding_conversion}
The multi-threshold function $f_{\Theta}(x)$ maps a real number $x$ to an integer in the interval $\left[0, N\right]$, which is the number of thresholds $\Theta_{i}$ that $x$ is greater than or equal to.
\ian{what is the number? N? later N is referred to as the number of steps. Maybe we can clarify here?}
At the output, a quantization scale $s_{\Theta}$ and bias $b_{\Theta}$ are applied, mapping the integers back to the real numbers.
The number of steps $N$ relates to the output bitwidth of the quantizer as $N = 2^{n} - 1$, where $n$ is the number of bits:
\begin{align}
    f_{\Theta}(x) = b_{\Theta} + s_{\Theta} \sum_{i=0}^{N-1}{\left(x \ge \Theta_{i}\right)}
    \label{eq:multi-threshold}
\end{align}
Following scale and bias aggregation, quantization scale and bias have already been extracted and pushed further down the computational graph, effectively setting $s_{\Theta}=1$ and $b_{\Theta}=b_{\text{sign}}$, which is the so-called sign-bias, shifting the output into the negative range in case of a signed quantizer, this also allows for narrow-range quantization (\autoref{sec:qonnx}):
\begin{align}
    b_{\text{sign}} = 1_{\text{signed?}}\left(-2^{n - 1} - 1 + 1_{\text{narrow?}}\right)
    \label{eq:sign-bias}
\end{align}

For integer inputs $x \in \mathbb{Z}$, such as those produced by an integer matrix multiplication (after scale-bias aggregation), rounding each threshold up to the next integer value does not change the output of the function
Similarly, for inputs constrained to a range $\left[\ii{x},\ia{x}\right]$, thresholds can be clipped to these bounds without changing the output of the function:
\begin{align}
    x \in \left[\ii{x},\ia{x}\right] \land x \in \mathbb{Z} \implies \sum_{i=0}^{N-1}{\left(x \ge \Theta_{i}\right)} = \sum_{i=0}^{N-1}{\left(x \ge \text{clip}\left(\lceil\Theta_{i}\rceil,\ii{x},\ia{x}+1\right)\right)}
    \label{eq:round-and-clip-thresholds}
\end{align}
This reduces the thresholding complexity based on the range information provided by SIRA or based on data type bounds of the input. For integer inputs, it is sufficient to represent thresholds as integers with as many bits as required to represent the input. The whole quantized layer tail reduces to an integer sum over integer comparisons.

\begin{figure}[!htbp]
    \centering
    \includegraphics[scale=0.8]{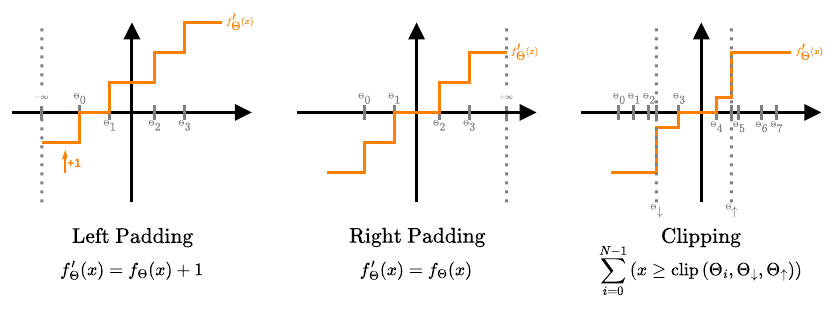}
    \caption{Padding and Clipping of Thresholds: Left padding thresholds by $-\infty$ has the effect of shifting the function upwards. Right padding by $+\infty$ does not affect the output of the function. Clipping the thresholds within $\left[\Theta_{\downarrow}, \Theta_{\uparrow}\right]$ pulls the function output outside this range to the upper/lower bound of the output range but does not affect the function output within the clipping range.}
    \label{fig:pad-and-clip-thresholds}
\end{figure}

Padding (inserting $-\infty$ from the left, or $+\infty$ from the right) and clipping thresholds according to \autoref{eq:round-and-clip-thresholds} are illustrated in \autoref{fig:pad-and-clip-thresholds}. Right padding of thresholds is used to fill the threshold list in case the conversion method does not find the expected number $N$ of thresholds. This does not affect the output of the function. Left padding has the same effect as adding a positive amount to the bias $b_{\Theta}$ and is used to offset the output without having to modify the bias. This comes in handy for realizing so-called stuck channels (\autoref{sec:results_stuck_channels}) (i.e., channels with constant output) and thus, without any actual thresholds. These padding thresholds are not really implemented as infinities---infinities are proxies, meaning \emph{any value outside of the input range}.%

\begin{figure}[!htbp]
  \centering
  \includegraphics[scale=0.8]{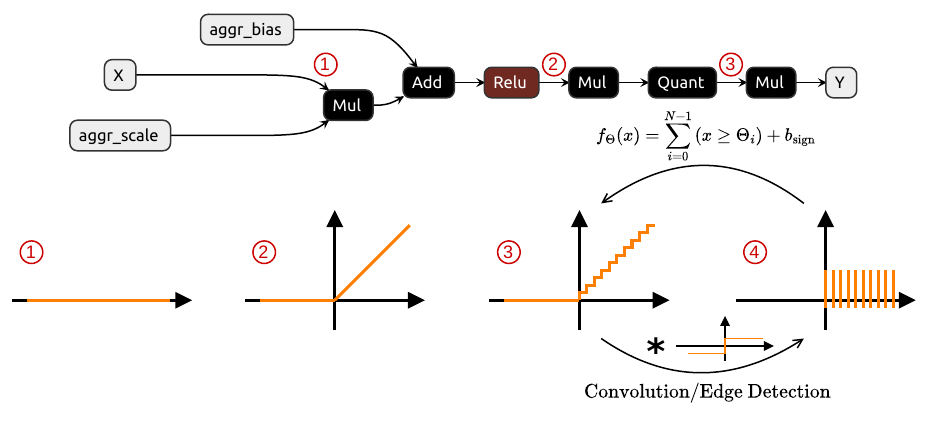}
  \caption{Converting Quantized Functions to Thresholds: Steps 1-3 represent the layer tail subgraph evaluation, starting from the SIRA-annotated input range (1), passing through scales, biases and the non-linearity, here ReLU (2), and terminating at the output quantizer (3), introducing the quantization steps. These steps are then picked up as the thresholds (4) by convolution with an edge detection kernel. From the extracted thresholds, the quantized layer tail output can be recovered according to Equation \eqref{eq:multi-threshold}.}
  \label{fig:threshold_conv}
\end{figure}

Instead of absorbing scales and biases by updating the thresholds step-by-step as proposed by \citet{umuroglu_jahre:CASES2017}, we convert the whole layer tail to a multi-threshold representation in a single step, based on information provided by SIRA: With the scale $s_{\Theta}$ and bias $b_{\Theta}$ known in advance, the entire behavior of $f_{\Theta}(x)$ is characterized by the thresholds~$\Theta_{i}$. These are the steps of a piecewise constant function, which can be extracted by evaluating the function over the whole range of possible inputs. This range is given by SIRA alongside the resolution of steps within this range. This perspective on the threshold conversion naturally extends to longer and more complex layer tails. Instead of formalizing local operator-specific rules for generating and updating the thresholds, a single formalization based on observing the end-to-end behavior of a subgraph is sufficient.
We implement this by convolving the function output of the quantized layer tail by an edge-detection kernel as shown in Figure \ref{fig:threshold_conv}.
Our proposed implementation is illustrated in \autoref{fig:threshold_conv}.
This not only yields the locations of the thresholds, but also the weight (i.e., height of the corresponding step), which is always a multiple of the quantization scale $s_{\Theta}$, or, after factoring this out during scale and bias aggregation, an integer.
Conceptually, this method works on any element-wise (no mixing across channels), non-broadcasting (broadcasting of parameter tensors is supported) layer tail that terminates in a uniform quantizer. Range information is needed at the input to the layer tail to have bounds for subgraph evaluation and threshold clipping.
The resulting threshold granularity of a layer tail is the finest granularity of any of the fused operators (e.g., if a per-tensor quantizer is preceded by \texttt{BatchNormalization}, the thresholds will show per-channel granularity picked up from the \texttt{BatchNormalization}).

The implementation builds upon existing ONNX and QONNX infrastructure for extracting and evaluating the subgraph, inserting the multi-threshold operation with potential quantization scale and bias, and removing nodes of the original subgraph.
The conversion pass starts at the end of the graph with the final quantizer, working upwards as we are always anchoring at the final quantizer of a layer tail to ensure fusing maximally extending subgraphs. Conceptual and implementation-specific restrictions are implemented as part of a layer tail subgraph extraction procedure. 
As our thresholding kernel implementation \autoref{sec:thresholding_hardware} is limited to handling positive unit steps and either per-tensor or per-channel granularity, layer tails with non-monotonic behavior or per-group granularity cannot be implemented.

\subsection{Accumulator Minimization With \OurScheme{}}
\label{sec:accumulator-sizing}

Accumulators are responsible for aggregating the partial results generated by multiplications.
For MAC-intensive layers like convolutions and matrix multiplication, the relative cost of accumulation grows progressively larger as inputs are quantized, and may even dominate overall MAC cost for few-bit quantization~\cite{colbert2024accumulator}.
\textit{Accumulator minimization} is the process of selecting the minimum bitwidth required for the accumulator, and can be used to control this cost.
The cost reduction is even more pronounced in \acrshort{fdna}s since datapath widths propagate from one layer to the next, the impact of accumulator width extends beyond the MAC layer, affecting the resource cost of subsequent elementwise layers and thresholding.
However, this requires careful consideration to avoid overflows, as reducing the accumulator widths below what the QNN requires can be catastrophic for ML task accuracy~\cite{colbert2023a2q}. 
In fixed-architecture hardware accelerators, an accumulation width of 32~bits or larger is typically used to avoid overflows.
A \emph{datatype bound}~\cite{colbert2023a2q} is a more common way to minimize accumulator size for \acrshort{fdna}s based on the maximum magnitudes of the input and weight datatypes.
Following the formulation by Colbert \textit{et al.}~\cite{colbert2023a2q}, assuming a $K$-dimensional dot product between $N$-bit unsigned inputs and $M$-bit signed weights, the minimum size of the datatype-bound accumulator can be computed as $P = \lceil \alpha + \phi(\alpha) + 1 \rceil$, with $\alpha = \mathrm{log}_2(K)+N+M-1$ and $\phi(\alpha) = \mathrm{log}_2(1+2^{-\alpha})$.

\begin{figure*}
  \centering
      \begin{subfigure}[b]{0.45\textwidth}
         \centering
         \includegraphics[width=\textwidth]{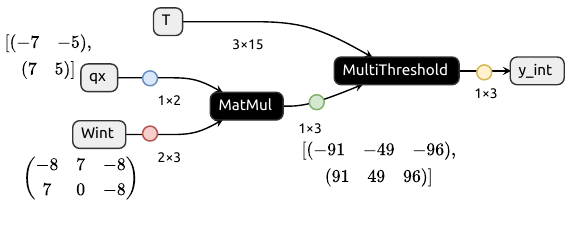}
         \caption{QNN subgraph with analysis results.}
         \label{fig:accmin-qnn}
     \end{subfigure}
     \hfill
     \begin{subfigure}[b]{0.54\textwidth}
         \centering
         \includegraphics[width=\textwidth]{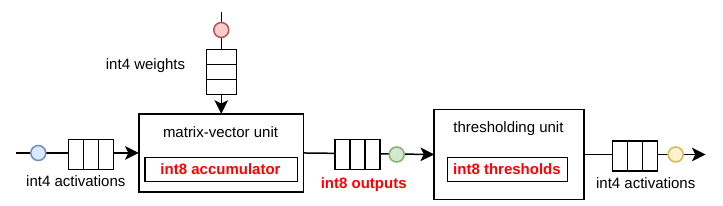}
         \caption{Corresponding hardware datapath.}
         \label{fig:accmin-hw}
     \end{subfigure}
  \caption{Minimizing the accumulator of a \texttt{MatMul} with \OurScheme{} and effect on subsequent \texttt{MultiThreshold} layer. Colored dots indicate corresponding tensors in the QNN and \acrshort{fdna} hardware. \textbf{\color{red}{Bold red}} indicates datapath widths affected by the optimization.}
  \label{fig:accumulator_sizing}
\end{figure*}

Owing to the guarantees provided by interval arithmetic in \OurScheme{}, we can perform further accumulator minimization for \acrshort{fdna}s without affecting the ML task accuracy.
After scales and biases are aggregated to reveal integer \texttt{Conv} and \texttt{MatMul} kernels, as explained in \autoref{sec:aggregate-scale-bias}, we can examine the integer ranges of these outputs to determine how many bits are needed to represent the output, which is sufficient for lossless accumulation.
Specifically, for a \texttt{Conv} or \texttt{MatMul} layer producing the signed integer output interval $\ivar{z}$, we can compute the required two's complement accumulator precision as $P = \lceil \log_2(\max(|\varmin{z}|, |\varmax{z}|+1)) \rceil + 1$, with extra +1 operations accounting for the reduced representational range for positive numbers and for the sign bit.
This exploits the case where weights are constant during inference and yields smaller accumulators than those calculated via the datatype bound.

\autoref{fig:accumulator_sizing} illustrates accumulator minimization for an example subgraph (\autoref{fig:accmin-qnn}) by showing the correspondence to the generated hardware (\autoref{fig:accmin-hw}). 
In this example, $P = \lceil \log_2(96+1) \rceil + 1 = 8$ bits are sufficient to accumulate the \texttt{MatMul} losslessly. 
By reducing the accumulator width, we also indirectly optimize the subsequent \texttt{MultiThreshold} layer, since each threshold parameter can now be made as small as the accumulator width.  
We note that this optimization is not limited to \texttt{MultiThreshold} layers; any subsequent layers which derive their bitwidths from the accumulator width, such as inter-layer and residual FIFOs, data width converters and elementwise layers for applying scales, biases and activation functions will similarly be reduced in terms of FPGA resource requirements.

\section{Compiler and Hardware Kernel Integration}
\label{sec:finn-integration}

\begin{figure}
  \centering
  \includegraphics[width=0.95\linewidth]{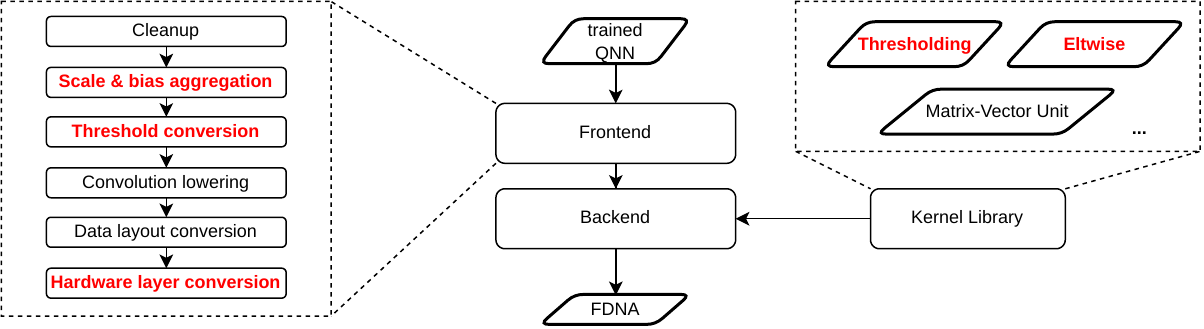}
  \caption{\OurScheme{} integrated into FINN compiler flow. New or modified components in \textbf{\color{red}{bold red}}.}
  \label{fig:finn-sira-flow}
\end{figure}

Compilers like FINN~\cite{umuroglu2017finn, blott2018finn} and hls4ml~\cite{duarte2018fast} generate \acrshort{fdna}s by first optimizing the graph before composing parametrized kernels instantiated from a hardware kernel library into a streaming dataflow pipeline.
In this section, we describe our extensions to FINN to produce \acrshort{fdna}s optimized by \OurScheme{}. 
This encompasses both novel graph optimization steps and new hardware kernels to realize the particular optimizations on the FPGA.
Although our implementation is FINN-based, these optimizations are portable to other \acrshort{fdna} compilers that leverage the QONNX representation.

\subsection{\OurScheme{}-enhanced FINN flow}

\begin{figure}
  \centering
  \includegraphics[width=0.95\linewidth]{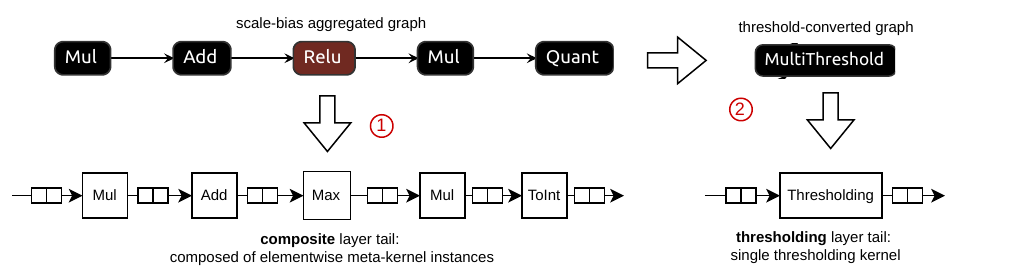}
  \caption{Implementation alternatives for layer tails in our enhanced FINN compiler flow.}
  \label{fig:tail-modes}
\end{figure}

The FINN compiler takes a trained quantized neural network in the QONNX format (\autoref{sec:qonnx}) as input. As illustrated in \autoref{fig:finn-sira-flow}, its \textit{frontend} optimizes and converts the QONNX graph into a format where each layer can be implemented by one or more hardware kernels, and its \textit{backend} configures, instantiates, and integrates hardware kernels with on-chip FIFO buffers in between to produce the FDNA.
The hardware kernels may be implemented in RTL or HLS, and support a variety of precision and parallelism options.
A seminal example is the Matrix-Vector Unit \cite{alam2023mvau}, which implements efficient MAC computations for convolutional and fully-connected layers.

Our extensions of the FINN compiler flow are highlighted in \autoref{fig:finn-sira-flow}.
Specifically, we integrate the \OurScheme{}-based streamlining optimizations of scale and bias aggregation (\autoref{sec:aggregate-scale-bias}) and extend and generalize threshold conversion (\autoref{fig:threshold_conv}) as new steps in the frontend.
Additionally, we (a)~extend the kernel library by the new RTL kernel for thresholding (\autoref{sec:thresholding_hardware}) and a new HLS meta-kernel for elementwise operations (\autoref{sec:eltwise_hardware}), and (b)~configure accumulation widths in MAC units based on \OurScheme{} (\autoref{sec:accumulator-sizing}).

The enhanced FINN flow offers two different modes for implementing scale-bias aggregated layer tails for \acrshort{fdna}s, as shown in \autoref{fig:tail-modes}.
One option is the \emph{composite layer tail} indicated by \circled{1} in the figure, where we directly convert each elementwise operator to a corresponding HLS meta-kernel.
The other option is the \emph{thresholding layer tail} indicated by \circled{2}, which first performs threshold conversion and implements this using the RTL thresholding kernel.
The layer tail mode is user-selectable, and our results in \autoref{sec:results-microbenchmarks} offer insights into the trade-offs between the two choices.

\subsection{Elementwise Operation Meta-Kernel}
\label{sec:eltwise_hardware}

\begin{figure}[htbp]
    \centering
    \includegraphics[scale=0.6]{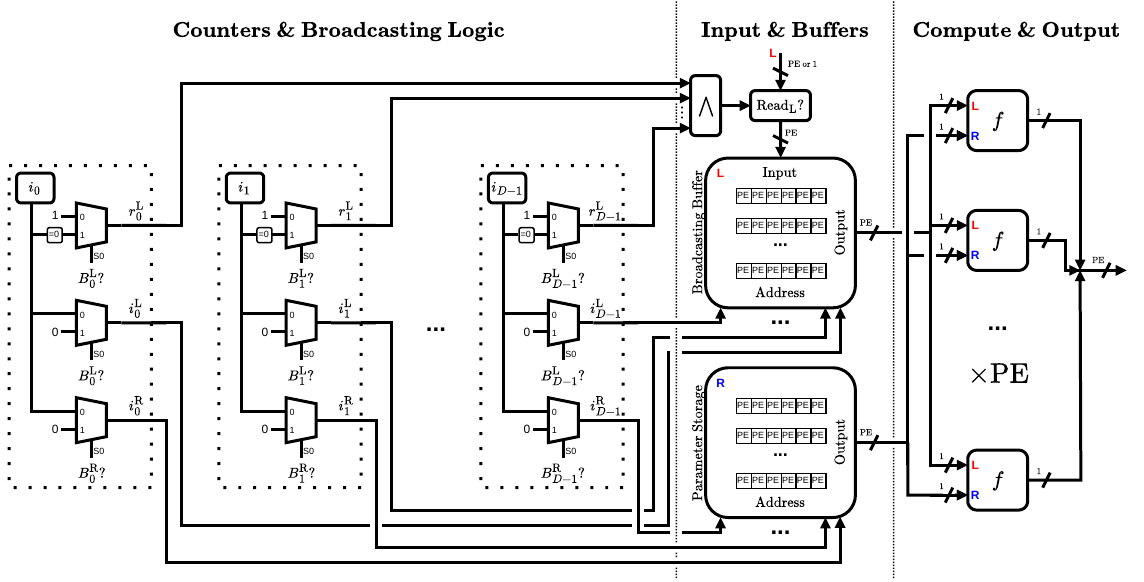}
    \caption{Elementwise Operation: The binary operation $f$ is applied to all elements of the input {\color{red}$\text{L}$} which connects to an input stream and the input {\color{blue}$\text{R}$} which is an embedded constant parameter storage. The function $f$ is replicated according to the $\text{PE}$-parallelism. The broadcasting logic computes buffer addresses from the fully broadcast indices %
    which reduces to $0$ if the dimension is broadcast for the respective input (i.e., $B^{\text{L}}_{d}?$ means ``if dimension $d$ of {\color{red}$L$ }is broadcast'') --- this can be computed in advance. The indices also contribute to the stream read condition. %
    Not shown here is the implementation of the counters $i_{0},i_{1},...,i_{D-1}$ realizing the loop-nest.
    }
    \label{fig:eltwise_hardware}
\end{figure}

To implement the individual operations of a composite layer tail, we extend and integrate an elementwise operation meta-kernel originally proposed for the FINN backend by \citet{berganski2024finnt}.
This is a Python-based C++ code-generation template describing a pipelined loop-nest for high-level synthesis, where the innermost loop is unrolled to parallelize across several output channels.
The degree of parallelization, denoted as PE (processing element) parallelism in FINN terminology \cite{blott2018finn, umuroglu2017finn}, is configured by the compiler according to the desired throughput.
Code-generation injects arbitrary elementwise (binary) operations, such as arithmetic, logical and bit-wise operations, or comparisons.
Each input can be either connected to an input stream or an embedded constant parameter storage.
The datatype support of these operators only relies on the capabilities of Vitis~HLS and comprises integer, floating-point (half, single, and double precision) and fixed-point formats.
See \autoref{fig:eltwise_hardware} for a schematic of such an elementwise operation.

Our implementation supports multidirectional broadcasting according to the ONNX standard.
This means, elementwise operations can operate on inputs of arbitrary shapes if they: 
(1) all have the same shape, 
(2) all have the same number of dimensions and the size of each dimensions is either a common size or 1, or, 
(3) can be left-padded with dimensions of size 1 to satisfy property 2.
Broadcasting is implemented by reading inputs into a broadcasting buffer ---
for constant parameters there is no extra buffer besides the parameter storage
--- from which they can be read out repeatedly.
This is achieved by generating an index $i_{d}$ for non-broadcast dimensions but reducing to a hard-coded $0$ for broadcast dimensions to reuse a single buffer slot.
The same principle applies to the buffer shapes.
During code generation, shape calculations align and pad the shapes, verify the broadcasting conditions and derive the output shape from broadcasting all inputs.
To minimize the size of the broadcasting buffer, reads from input streams are delayed until the elements are needed according to broadcasting semantics.
A broadcasting-read condition is derived for each input. %

The loop-nest template injects compiler directives to guide optimization of the design: The operand memories are partitioned along the last dimension to enable the PE-parallel access pattern of the unrolled innermost loop. These memories are set to be implemented as dual-port RAMs (broadcasting buffers) or ROMs (constant parameter storage) and can optionally be forced to use either LUT, BRAM, or URAM resources. Similarly, some arithmetic operations, such as multiplications and additions, can be forced to either LUT or DSP-based implementation.

\subsection{Thresholding Kernel}
\label{sec:thresholding_hardware}

\begin{figure}[htb]
    \centering
    \includegraphics[scale=0.6]{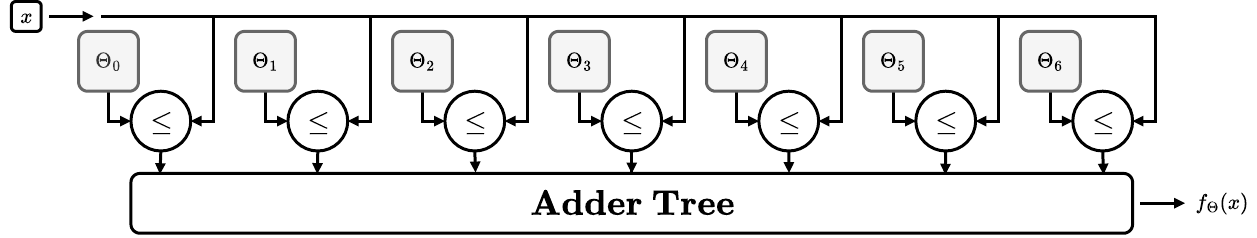}
    \caption{Thresholding by parallel comparators and summation illustrated for a 3-bit result $f_{\Theta}(x)$ na\"ively realizing \autoref{eq:multi-threshold} for a set of seven thresholds. This requires seven comparators followed by an adder tree.}
    \label{fig:parallel_thresholding}
\end{figure}

A hardware kernel for the multi-threshold operator (\autoref{sec:thresholding_conversion}) has existed in FINN since the original proposals in~\cite{umuroglu2017finn, blott2018finn}, as further detailed in \autoref{sec:related-finn-thresholding}.
This previous implementation, as a straightforward parallel comparison with all thresholds and a summation, illustrated in \autoref{fig:parallel_thresholding}, has several drawbacks.
As the number $N$ of thresholds grows exponentially with the bitwidth $n$ of the produced output, scaling this approach beyond $n=4$ bits becomes notably costly in terms of threshold parameter bandwidth and comparator resources.

\begin{figure}
    \centering
    \includegraphics[scale=0.6]{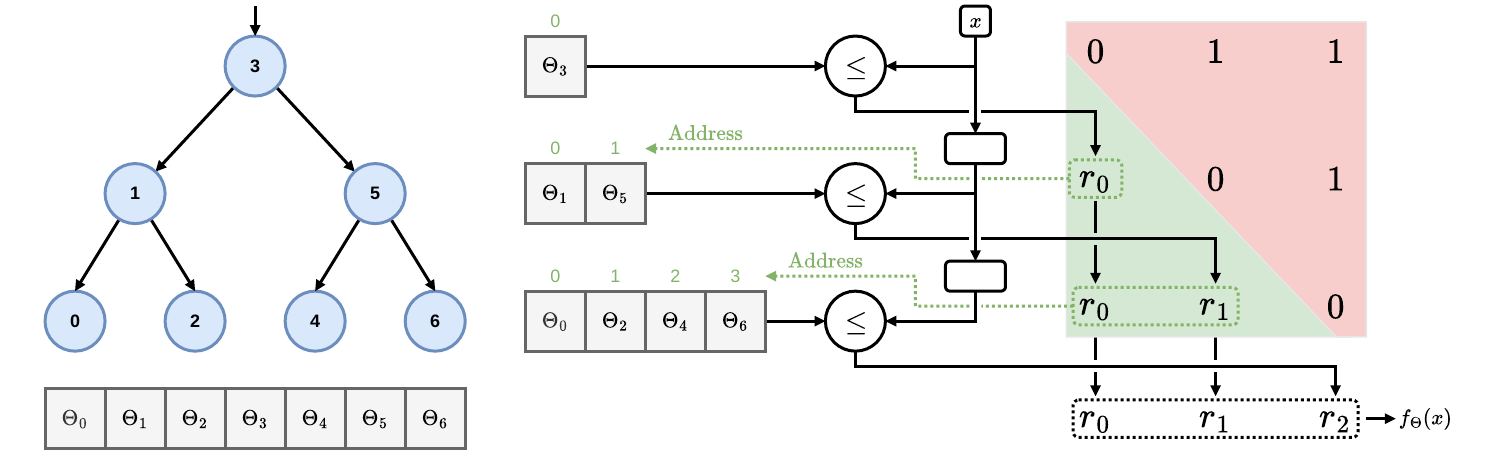}
    \caption{Thresholding by binary search illustrated for a 3-bit result $f_{\Theta}(x)$ determined by locating the input $x$ within a \emph{sorted} array of seven thresholds organized as a binary search tree as shown on the left side. The right side shows the corresponding pipelined traversal of this tree, requiring only three comparators compared to the seven of the parallel implementation shown in \autoref{fig:parallel_thresholding}.}
    \label{fig:binary_search_thresholding}
\end{figure}

\yaman{should we drop the naive parallel comparator figure and shorten its description to save space, since we don't provide any results to compare against it? } \christoph{If really in need of saving more space, yes... but I like this from a story-telling perspective, where this kind of sits in the middle, bridging the gap between \autoref{eq:multi-threshold} and \autoref{fig:binary_search_thresholding}. I'd rather drop \autoref{fig:qonnx_ecosystem} and/or \autoref{lst:node-by-node} if necessary to save more space.}

To enable a more efficient thresholding operation,
we extend the kernel library by an alternative thresholding implementation based on a binary search within the \emph{sorted} array of threshold values.
It unrolls the steps of the search into a structural pipeline.
Each pipeline stage advances the traversal of the binary search tree by one level.
While the first stage always splits the input domain by the median threshold, the subsequent ones use the concatenated preceding comparison outcomes to select the increasingly fine-grained decision that needs to be made next.
The process is illustrated for a small problem size of $n=3$ output bits in \autoref{fig:binary_search_thresholding}. 
Note how each pipeline stage relies on its own local storage for holding all thresholds of the level of the binary search tree that it represents. This is all that is needed for making the next search decision.

The local addresses of thresholds are extended into global threshold indices as illustrated by the highlighted red triangular matrix completion.
The scheme can be understood by observing that the last level of a complete search tree comprises only and all its nodes with even indices.
This property holds recursively for the truncated address space of shallower subtrees. Thus, all global address extensions follow the pattern comprising a single zero bit followed by a sequence of ones, padding it all the way up to the global address size. Underutilized output coding spaces that differentiate fewer than $2^n$ values and, hence, require fewer than $2^n-1$ thresholds are facilitated by extending the threshold array by values explicitly identified as infinite for the purpose of comparison. This ensures that the corresponding decision paths are never traversed and the highest values in the output coding space will never be produced. An optional \texttt{BIAS} parameter can be used to shift the output interval, for example, to implement the sign bias introduced in \autoref{sec:thresholding_conversion}.

\subsection{Analytical Models}
\label{sec:analytical_cost_models}

To help determine whether composite or thresholding layer tails are advantageous for a particular QNN without FPGA  synthesis, we provide analytical cost models below.

\subsubsection{Elementwise Operation Meta-Kernel}
\label{sec:eltwise_kernel}

We consider each type of common layer tail operation individually, focusing on fixed-point data types for practical layer tail implementations (\autoref{sec:fixed-pt-quant}), and enforce a LUT-based implementation to facilitate direct comparisons to the analytical model for thresholding layers.
Each operation type has a different correlation with the input bitwidth $n_i$ and with the parameter bitwidth $n_p$. 
We assume that \texttt{Mul} nodes have a multiplicative relationship between $n_i$ and $n_p$ for the predicted resource cost, while the remaining nodes have a linear relationship.
We multiply by the number of processing elements ($\text{PE}$), as explained in \autoref{sec:eltwise_hardware}, before calculating each $\alpha$ and $\beta$ in \autoref{tab:eltwise-analutical} via linear regression over out-of-context synthesis results for a range of fixed-point data types. 
\autoref{fig:eltwise_resource_cost_evaluation} plots the predictions against observed synthesis results across a range of configurations of $n_i$, $n_p$, $\text{PE}$, and supported operations, indicating a mean relative error of $4\%$.

\begin{center}
  \centering
  \begin{minipage}[c]{0.45\textwidth}
    \centering
%

\definecolor{mycolor1}{rgb}{0.00000,0.44700,0.74100}%
\definecolor{mycolor2}{rgb}{0.85000,0.32500,0.09800}%
\definecolor{mycolor3}{rgb}{0.92900,0.69400,0.12500}%
\definecolor{mycolor4}{rgb}{0.49400,0.18400,0.55600}%
\definecolor{mycolor5}{rgb}{0.46600,0.67400,0.18800}%
\definecolor{mycolor6}{rgb}{0.30100,0.74500,0.93300}%
\definecolor{mycolor7}{rgb}{0.63500,0.07800,0.18400}%

\begin{tikzpicture}[%
trim axis left, trim axis right
]

\begin{axis}[%
height=0.60\columnwidth,
width=0.90\columnwidth,
xmin=0.00,
xlabel style={font=\color{white!15!black}},
xlabel={Configuration ($n_i$, $n_p$ $PE$)},
xlabel shift=-3pt,
ymin=0.00,
ymax=10000.00,
ylabel style={font=\color{white!15!black}},
ylabel={$\text{LUT}$},
axis background/.style={fill=white},
axis lines = left,
ymajorgrids,
yminorgrids,
ymode=log,
log basis y=10, 
log ticks with fixed point,
legend style={at={(0.5,1.12)},anchor=center,font=\footnotesize},/pgf/number format/1000 sep={},
legend columns=2,
align=left,
tick label style={/pgf/number format/assume math mode = true},
xlabel style={font=\normalsize},
ylabel style={font=\normalsize},
x tick label style={/pgf/number format/.cd, fixed, fixed zerofill, precision=0, /tikz/.cd, font=\small},
y tick label style={ /pgf/number format/.cd, fixed, fixed zerofill, precision=0, /tikz/.cd, font=\small},
enlargelimits =false,clip=false,
]

\addplot [mark=*, only marks, mark size=0.08cm, mark options={fill=white, draw=mycolor1}] table [ color=mycolor1, x=x, y=y_predicted_luts_max, col sep=comma] {figures/graphs/eltwise.csv};
\addlegendentry{Predicted Max}

\addplot [mark=*, only marks, mark size=0.03cm, color=mycolor1] table [ color=mycolor1, x=x, y=y_actual_luts_max, col sep=comma] {figures/graphs/eltwise.csv};
\addlegendentry{Actual Max}

\addplot [mark=*,only marks, mark size=0.08cm, mark options={fill=white, draw=mycolor2}] table [ color=mycolor1, x=x, y=y_predicted_luts_add, col sep=comma] {figures/graphs/eltwise.csv};
\addlegendentry{Predicted Add}

\addplot [mark=*, only marks, mark size=0.03cm, color=mycolor2] table [ color=mycolor1, x=x, y=y_actual_luts_add, col sep=comma] {figures/graphs/eltwise.csv};
\addlegendentry{Actual Add}

\addplot [mark=*, only marks, mark size=0.08cm, mark options={fill=white, draw=mycolor3}] table [ color=mycolor1, x=x, y=y_predicted_luts_round, col sep=comma] {figures/graphs/eltwise.csv};
\addlegendentry{Predicted ToInt}

\addplot [mark=*, only marks , mark size=0.03cm, color=mycolor3] table [ color=mycolor1, x=x, y=y_actual_luts_round, col sep=comma] {figures/graphs/eltwise.csv};
\addlegendentry{Actual ToInt}

\addplot [mark=*, only marks, mark size=0.08cm, mark options={fill=white, draw=mycolor4}] table [ color=mycolor1, x=x, y=y_predicted_luts_mul, col sep=comma] {figures/graphs/eltwise.csv};
\addlegendentry{Predicted Mul}

\addplot [mark=*, only marks , mark size=0.03cm, mark options={fill=mycolor4, draw=mycolor4} ] table [ color=mycolor1, x=x, y=y_actual_luts_mul, col sep=comma] {figures/graphs/eltwise.csv};
\addlegendentry{Actual Mul}

\end{axis}

\end{tikzpicture}%
    \captionof{figure}{Analytical cost model of elementwise operations.}
    \label{fig:eltwise_resource_cost_evaluation}
  \end{minipage}
  \hfill
  \begin{minipage}[c]{0.49\textwidth}
    \captionof{table}{Analytical cost model of the elementwise operation meta-kernel. PE: processing element, $n_i, n_p$: bitwidth of inputs.}
    \label{tab:eltwise-analutical}
    \small
    \begin{tabular}{cccc}
    \textbf{Operation} & \textbf{LUTs} & $\mathbf{\alpha}$ & $\mathbf{\beta}$ \\ \hline
    \multicolumn{1}{c|}{\texttt{Mul}} & \multicolumn{1}{c|}{$\alpha*n_i*n_p*\text{PE}+\beta$} & \multicolumn{1}{c|}{1.18} & 124 \\
    \multicolumn{1}{c|}{\texttt{Add}} & \multicolumn{1}{c|}{$\alpha*(n_i+n_p)*\text{PE}+\beta$} & \multicolumn{1}{c|}{2.0} & 24 \\
    \multicolumn{1}{c|}{\texttt{ToInt}} & \multicolumn{1}{c|}{$\alpha*n_i*\text{PE}+\beta$} &  \multicolumn{1}{c|}{4.2} & 13 \\
    \multicolumn{1}{c|}{\texttt{Max}} & \multicolumn{1}{c|}{$\alpha*n_i*\text{PE}+\beta$} & \multicolumn{1}{c|}{4.0} & 21
    \end{tabular}
  \end{minipage}
\end{center}
\felix{There is still a minor LaTeX issue here where the clickable references (hypcap package) to these figures/tables jump to the caption instead of the top of the figure.}

\subsubsection{Composite Layer Tail}
\label{sec:eltwise_cost_model}

We model the composite layer tail as consisting of 5 nodes after scale-bias aggregation as illustrated in \autoref{fig:tail-modes}.
Based on ~\autoref{tab:eltwise-analutical}, we calculate the LUT usage for computation for each of the operations as a function of $n_i$ and $n_p$ (when present).
We also account for the requirements of lossless fixed-point arithmetic within the layer tail. 
When multiplying two fixed-point numbers with different bitwidths, the resulting output bitwidth is equal to the sum of the input bitwidths. 
In the case of addition, the output bitwidth must accommodate the larger of the two input bitwidths, plus one additional bit to handle potential overflow. 
This results in $\text{LUT}_{\text{comp}}(n_i,n_p,\text{PE})  = \text{Mul}(n_i,n_p,\text{PE}) + \text{Add}(n_i+n_p,n_p,\text{PE})+\text{Max}(n_i+n_p+1,\text{PE})+ \text{Mul}(n_i+n_p+1,n_p,\text{PE}) + \text{ToInt}(n_i+n_p+1,\text{PE}) )$.

To account for the memory cost of layer tail parameters, we compute the number of LUTs that would be used as memory. 
With per-channel weight and per-tensor activation scales (\autoref{sec:sira-matmul-and-conv}), only two operations (\texttt{Mul}, \texttt{Add}) have per-channel parameters.
Thus, for $C$ channels and 6-input LUTs, we model $\mathrm{LUT}_\mathrm{mem}(n_p, C) = 2 \cdot C \cdot \frac{n_p}{64}$.
The final predicted amount of LUTs can then be calculated by adding the contributions of computation and memory, so $\text{LUT}_{\text{total}}(n_i,n_p,C,\text{PE}) = \text{LUT}_{\text{comp}}(n_i,n_p,\text{PE}) + \mathrm{LUT}_\mathrm{mem}(n_p, C)$.

\subsubsection{Thresholding Kernel}
\label{sec:multi_thres_cost_model}

To simplify the creation of the analytical model, we differentiate between FPGA resources used for memory and those used for computation.  
The memory contribution is based on the number of thresholds and their own bitwidth. 
The sum of thresholds $\mathrm{Sum}_{\mathrm{\Theta}}$ depends on the output bitwidth $n_o$ of the operation and the number of channels $C$ as $\mathrm{Sum}_{\Theta} = (2^{n_o} - 1) * C$. 
The number of channels is determined by the granularity of the weight scale factor and by the other operations in the layer tail that have been fused into the multi-threshold function as described in \autoref{sec:thresholding_conversion}. 
Each threshold parameter has the same bitwidth $n_i$ as the input bitwidth. %
Thus, the memory expressed in terms of bits is calculated as $\mathrm{MEM}_{\mathrm{bits}} = \mathrm{Sum}_{\mathrm{\Theta}} * n_i$.
Depending on the available resources, %
the thresholds can be stored in either LUTs or BRAMs. 
For easier comparison, we consider only 6-input LUTs, thus, $\mathrm{LUT}_{\mathrm{mem}} =
       \frac{\mathrm{MEM}_{\mathrm{bits}}} {64}$ 
       
The computation contribution is based on the number of bit comparators present in the thresholding kernel. 
The number of comparators calculated will correspond directly to the number of processing elements and the output bits, while the bit comparators are based on the number of comparators and the input bitwidth as $\mathrm{LUT}_{\mathrm{comp}} = n_o * \mathrm{PE} * n_i$. PE refers to the parallelism as introduced in \autoref{sec:eltwise_hardware}.
The final prediction is calculated by adding the memory and computation contributions, so $\text{LUT}_{\text{total}}(n_i,n_p,C,\text{PE},n_o) = \text{LUT}_{\text{comp}}(n_i,n_o,\text{PE}) + \mathrm{LUT}_\mathrm{mem}(n_i,n_o,C)$

\begin{figure}[!htbp]
\centering
%

\definecolor{mycolor1}{rgb}{0.00000,0.44700,0.74100}%
\definecolor{mycolor2}{rgb}{0.85000,0.32500,0.09800}%
\definecolor{mycolor3}{rgb}{0.92900,0.69400,0.12500}%
\definecolor{mycolor4}{rgb}{0.49400,0.18400,0.55600}%
\definecolor{mycolor5}{rgb}{0.46600,0.67400,0.18800}%
\definecolor{mycolor6}{rgb}{0.30100,0.74500,0.93300}%
\definecolor{mycolor7}{rgb}{0.63500,0.07800,0.18400}%

\begin{tikzpicture}[%
trim axis left, trim axis right
]

\begin{axis}[%
height=3cm,
width=0.95\columnwidth,
xlabel style={font=\color{white!15!black}},
xlabel={Configuration ($n_i$, $n_o$, $chan$, $PE$)},
xtick distance={20},
xlabel shift=-3pt,
ylabel style={font=\color{white!15!black}},
ylabel={$\text{LUT}$},
axis background/.style={fill=white},
axis lines = left,
ymajorgrids,
yminorgrids,
ymode=log,
log basis y=10, 
log ticks with fixed point,
legend style={legend cell align=left, align=left, draw=white!15!black},
legend style={at={(0.5,1.12)},anchor=center,font=\footnotesize},/pgf/number format/1000 sep={},
legend columns=2,
tick label style={/pgf/number format/assume math mode = true},
xlabel style={font=\normalsize},
ylabel style={font=\normalsize},
x tick label style={ /pgf/number format/.cd, fixed, fixed zerofill, precision=0, /tikz/.cd, font=\small},
y tick label style={ /pgf/number format/.cd, fixed, fixed zerofill, precision=0, /tikz/.cd, font=\small},
enlargelimits =false,clip=false,
]

\addplot [mark=*, only marks, mark size=0.05cm, color=mycolor1] table [ color=mycolor1, x=x+AF8-axis, y=actual, col sep=comma] {figures/graphs/thresholds.csv};
\addlegendentry{actual LUT}

\addplot [mark=*, only marks, mark size=0.05cm, color=mycolor2] table [color=mycolor2, x=x+AF8-axis, y=predicted+AF8-lut, col sep=comma] {figures/graphs/thresholds.csv};
\addlegendentry{predicted LUT}
\end{axis}

\end{tikzpicture}%
\caption{Analytical cost model of the multi-threshold operator when targeting LUT-only implementation. Sweep parameters: $n_i=[8,16,32]$, $n_o=[2,4,8]$, $\mathrm{chan}=[1,64,128,256,512]$, $\mathrm{PE}=[1,2,4]$.  Mean Relative Error (MRE) is $15\%$. }
\label{fig:resource_cost_multithresholds_lut_only}
\end{figure}
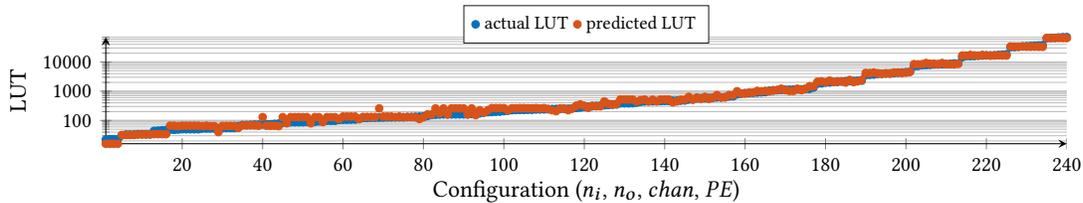

To evaluate the effectiveness of the analytical model, we perform out-of-context synthesis across a wide range of operator configurations.
We consider input bitwidths from 8 to 32 bits, output bitwidths from 2 to 8, channel ranges from 1 to 512, and PE from 1 to 4 \felix{do we have a reason behind this number? E.g., typical for our use cases, doesn't change much when moving past 4?}, making up a total of 244 configurations.
We cover all possible combinations across these parameters, targeting a frequency of 200 MHz with LUT-only implementation. 
\autoref{fig:resource_cost_multithresholds_lut_only} compares the measured to the predicted LUT utilization from the analytical model.

\section{Methodology}
\label{sec:methodology}

We describe the verification and evaluation methodology for \OurScheme{}. 
To assess the impact of \OurScheme{} optimizations for streamlining and accumulator minimization on \acrshort{fdna}s, we use FINN integrated with \OurScheme{} as described in \autoref{sec:finn-integration}, and evaluate several QNN workloads as well as microbenchmarks as described below.

\subsection{Verification}

In addition to unit tests for the propagation rules in \autoref{sec:sira-propaget} and threshold conversion in \autoref{sec:streamlining}, we perform empirical verification for ranges generated by our \OurScheme{} implementation by
comparing against instrumentation data. 
Instrumentation data are obtained by running inference on a trained neural network for each item in a dataset and tracking the minimums and maximums for intermediate tensors.

\subsection{QNN Workloads}
\label{sec:methodology-end2end}
\begin{table}
    \centering
    \caption{QNNs from QONNX model zoo used for end-to-end evaluation of \OurScheme{}.}
    \small
    \begin{tabular}{cccccccc}
    \toprule
         \multirow{2}{*}{Name} &   \multirow{2}{*}{Topology}& \multirow{2}{*}{Properties\textsuperscript{*}} &  \multirow{2}{*}{MACs}& \multirow{2}{*}{Parameters} & \multirow{2}{*}{Task} & \multicolumn{2}{c}{Top-1 Accuracy \%}\\
         &   &  & & & & Original & Fixed-point\\
     \midrule
 TFC-w2a2 & 3-layer MLP & f & 59k & 59k & MNIST & 96.60 & 96.46 \\
 CNV-w2a2 & VGG10-like & c, f & 60M & 1.5M & CIFAR-10 & 88.84 & 87.94 \\
 RN8-w3a3 & ResNet-8 & c, 8, r & 195M & 1.3M & CIFAR-100 & 70.13 & 69.16 \\
 MNv1-w4a4 & MobileNet-v1 & c, d, 8 & 568M & 4.2M & ImageNet-1k & 69.95 & 68.46 \\
    \midrule
    \multicolumn{8}{c}{\textsuperscript{*}(f)ully-connected, (c)onvolution, (d)epthwise convolution, (8)-bit first/last layer, (r)esidual connections}  \\
     \bottomrule
    \end{tabular}
    \label{tab:end2end_workloads}
\end{table}

We evaluate \OurScheme{} optimizations on image classification workloads from the QONNX model zoo~\cite{qonnx-model-zoo}.
As shown in \autoref{tab:end2end_workloads}, these workloads represent a range of topologies, model complexities, and quantization precision.
The workload names encode their predominant quantization as "wXaY' indicating X-bit weights and Y-bit activations.
All workloads use \texttt{BatchNormalization} layers before ReLU activation, which is equivalent to per-channel floating-point scaling for weights.
Activations use per-tensor floating-point scaling except when the following layer is a depthwise convolution, in which case per-channel floating-point scaling is used.
Other important aspects of the workloads, listed in the Properties column, cover common QNN vision building blocks and test \OurScheme{}'s applicability.
To separate the effect of the threshold conversion and accumulator minimization optimizations, we run a total of four synthesis experiments for each QNN with each optimization enabled or disabled.
When thresholding is disabled, we use the composite layer tail implementation mode (\autoref{fig:tail-modes}).
To avoid an inflated baseline with float-scaled MAC operations, we perform scale and bias aggregation (\autoref{sec:aggregate-scale-bias}) for all configurations, including the baseline.

\subsubsection{Fixed-point Quantization}
\label{sec:fixed-pt-quant}

When implementing composite layer tails as part of QNN workloads, we quantize the aggregated scale and bias values to fixed-point formats since floating-point computations are highly resource-intensive on our target FPGA. 
Although not part of the \OurScheme{} optimizations, this is a common approach for efficient inference \cite{duarte2018fast,  jacob2018quantization}, so we adopt it as part of the baseline.
To determine the fixed-point format \texttt{fixedW.I} for each tensor, we pick the number of integer bits \texttt{I} for lossless representation of the integer part of the corresponding floating-point range, and we pick the fractional bits \texttt{F} using grid search where \texttt{W=F+I}; the resulting accuracy drop compared to its floating-point counterpart is at most 1.5 percentage points.
This additional fixed-point quantization is only necessary for the experiments with composite layer tails as the thresholding does not require additional quantization.

\subsubsection{Folding Configuration}

For each QNN, we pick a throughput target that results in a layer pipeline without major imbalance while maximizing the throughput, subject to the 8,192-bit limit on Vitis HLS arbitrary-precision integer (\texttt{ap\_int}) types. 
Since FINN packs elements to be processed in parallel into \texttt{ap\_int} to carry them between layers, the output of an individual layer cannot be wider than this limit, thus limiting the available parallelism.

\subsection{Layer Tail Analysis}
\label{sec:methodology_layer_tail_analysis}

To gain a deeper understanding of the trade-offs between thresholding and composite layer tails, we generate a series of microbenchmarks corresponding to variants of \autoref{fig:tail-modes}.
These are implemented as either composite or thresholding layer tails, which are our main point of comparison for these benchmarks. 
Other configuration options cover the input and activation bitwidth, the granularity of scales and biases (per-tensor or per-channel), optional restriction of scales and biases to \ac{pot}, and whether to implement elementwise operators in \texttt{float32} or fixed-point \texttt{fixed16.8, fixed32.16} arithmetic. 
The number of channels and the parallelism are fixed at 256 and PE=4, respectively. 
Each number reported is the average across three independent synthesis runs, varying the random seed used to generate the model parameters.
Finally, we use the composite layer tail analytical model derived in \autoref{sec:analytical_cost_models} to project layer tail costs for a broader variety of configurations and understand the crossover point between composite and thresholding layer tails.

\subsection{FPGA Implementation}

We rely on FINN for FPGA implementation, which uses AMD Vitis HLS and SystemVerilog RTL kernels to instantiate \acrshort{fdna}s with different configurations.
To report FPGA resource results, we run out-of-context synthesis using Vivado 2024.2 with the default synthesis strategy, targeting a Zynq UltraScale+ XCZU9EG FPGA device.

\subsubsection{Memory and Computational Resources.}
Modern FPGAs have multiple types of memory and computational resources, which can be utilized to create \acrshort{fdna}s with different resource mixes. 
In our evaluation, we use the default \texttt{auto} memory resource setting to let Vitis HLS and Vivado decide the resource allocation for individual layer memories.
In terms of computational resources, to take advantage of the high-density MAC operations offered by FPGA DSP blocks for particular precisions, we use the FINN RTL MVU with DSP packing optimizations for 4-bit and 8-bit arithmetic, while MACs with other precisions are instantiated with LUTs by Vitis HLS.
For elementwise operation meta-kernels, we use the default Vitis HLS resource strategy, which may prefer DSPs, LUTs, or a combination of both.
The exceptions are the microbenchmark for comparing composite and thresholding layer tails, where we force all components use only LUTs for both computation and memory to enable easier analytical modeling and to make direct comparisons possible.

\subsubsection{Timing}

As our goal is to highlight relative resource improvements enabled by our optimizations, we do not focus on achieving high-frequency timing closure and only report out-of-context synthesis results. 
However, our on-board validation on a ZCU102 indicates that 200~MHz is achievable for the \OurScheme{}-optimized end-to-end benchmarks we study, so we pick this as the clock frequency for reporting throughput and latency numbers.
While we note that further resource reduction by full place-and-route with physical synthesis optimizations and performance increase by double-pumping the clocks for DSP slices are possible, we leave their further study for future work.

\section{Results}
\label{sec:results}
We present the results of the \OurScheme{} verification, our evaluation on \ac{qnn} workloads and layer tail microbenchmarks.

\subsection{Verification and Stuck Channels}
\label{sec:results_stuck_channels}
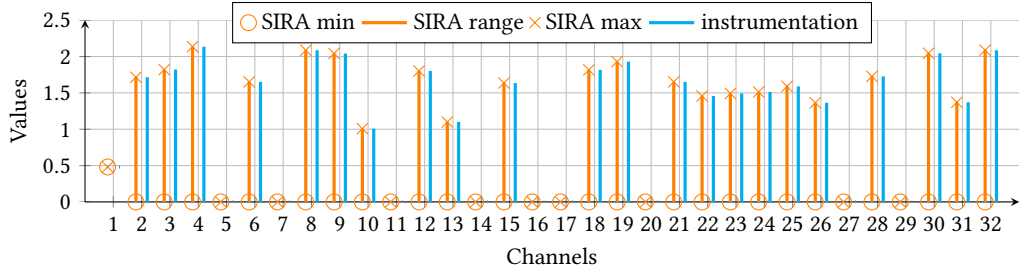
\begin{figure}
    \centering
\begin{tikzpicture}
\begin{axis}[
    width=14cm, height=4cm,
    ymin=0, ymax=2.5,
    xmin=0, xmax=33,
    xlabel={Channels},
    ylabel={Values},
    xtick={1,2,...,32},
    ytick distance=0.5,
    grid=both,
    legend columns=4,
    legend style={at={(0.5,1.1)}, anchor=north},
    legend entries={\OurScheme{} min, \OurScheme{} range, \OurScheme{} max, instrumentation},
    axis lines=left
]

\foreach \x/\y/\z in {
1/0.4813263/0.4813263, 2/0/1.715205, 3/0/1.8205122, 4/0/2.135432, 5/0/0, 6/0/1.6516743, 7/0/0, 8/0/2.0862162, 9/0/2.04194, 10/0/1.0084338, 11/0/0, 12/0/1.8011423, 13/0/1.1001043, 14/0/0, 15/0/1.6353799, 16/0/0, 17/0/0, 18/0/1.815996, 19/0/1.9297462, 20/0/0, 21/0/1.6514481, 22/0/1.4564672, 23/0/1.4921143, 24/0/1.5131319, 25/0/1.5893719, 26/0/1.3634044, 27/0/0, 28/0/1.7266119, 29/0/0, 30/0/2.045092, 31/0/1.3708094, 32/0/2.0866995
}
{
    \addplot+[only marks, mark=o, mark size=3pt, orange, forget plot] coordinates {(\x-0.2,\y)};
    \addplot+[no markers, orange, very thick, forget plot] coordinates {(\x-0.2,\y) (\x-0.2,\z)};
    \addplot+[only marks, mark=x, mark size=3pt, orange, forget plot] coordinates {(\x-0.2,\z)};
}

\foreach \x/\y/\z in {
1/0.48132631182670593/0.48132631182670593, 2/0.0/1.7152049541473389, 3/0.0/1.8205121755599976, 4/0.0/2.135432004928589, 5/0.0/0.0, 6/0.0/1.6516742706298828, 7/0.0/0.0, 8/0.0/2.0862162113189697, 9/0.0/2.0419399738311768, 10/0.0/1.0084338188171387, 11/0.0/0.0, 12/0.0/1.8011423349380493, 13/0.0/1.1001043319702148, 14/0.0/0.0, 15/0.0/1.6353799104690552, 16/0.0/0.0, 17/0.0/0.0, 18/0.0/1.8159960508346558, 19/0.0/1.929746150970459, 20/0.0/0.0, 21/0.0/1.651448130607605, 22/0.0/1.4564671516418457, 23/0.0/1.4921143054962158, 24/0.0/1.513131856918335, 25/0.0/1.589371919631958, 26/0.0/1.363404393196106, 27/0.0/0.0, 28/0.0/1.726611852645874, 29/0.0/0.0, 30/0.0/2.0450921058654785, 31/0.0/1.3708094358444214, 32/0.0/2.0866994857788086
}
{
    \addplot+[no markers, cyan, very thick, forget plot] coordinates {(\x+0.2,\y) (\x+0.2,\z)};
}

\addplot+[only marks, mark=o, mark size=3pt, orange] coordinates {(NaN,NaN)};
\addlegendimage{only marks, mark=o, mark size=3pt, orange}
\addplot+[no markers, orange, very thick] coordinates {(NaN,NaN)};
\addlegendimage{no markers, orange, very thick}
\addplot+[only marks, mark=x, mark size=3pt, orange] coordinates {(NaN,NaN)};
\addlegendimage{only marks, mark=x, mark size=3pt, orange}
\addplot+[no markers, cyan, very thick] coordinates {(NaN,NaN)};
\addlegendimage{no markers, cyan, very thick}

\end{axis}
\end{tikzpicture}
  \caption{Per-channel observed ranges from instrumentation and \OurScheme{} analysis for the first quantized activation layer in MNv1-w4a4.}
  \label{fig:verify_mnv1}
\end{figure}

\autoref{fig:verify_mnv1} plots the per-channel observed ranges on the ImageNet validation dataset and \OurScheme{} results on MNv1-w4a4, where all observations from instrumentation fall within the \OurScheme{}-reported range.
In general, the ranges reported by \OurScheme{} are correct, but conservative; the observed ranges always fall within the analysis ranges, but sometimes the analysis range is larger.
This is expected, since we cannot guarantee that the instrumentation process can observe all possible values without instrumenting all possible inputs, an infeasible task even for the smallest NNs.

We make another interesting observation based on \OurScheme{} analysis results: there are several \emph{stuck channels} in our QNN workloads, which always produce a constant (point interval) no matter their input.
In the example in \autoref{fig:verify_mnv1}, channel 1 produces a constant value of 0.48, while 9 other channels produce a constant value of 0, which can be seen as a generalization of the \emph{dying ReLU} problem~\cite{lu2019dying}.
As these channels offer no predictive power, they can be removed to reduce inference cost with no detriment to accuracy.
We leave further study of these stuck channels to future work.

\subsection{End-to-end QNN Workloads}
\label{sec:results-end2end}

\begin{table}[!htp]\centering
\caption{Out-of-context synthesis results for QNN workloads, with and without \OurScheme{} optimizations. 
\hms{Suggest to use \checkmark instead of + and then skip using -. Should make it easier to read.}
(\checkmark) in Thr and Acc columns indicate whether threshold conversion and accumulator minimization were applied.
The columns prefixed with r indicate the resources normalized to the non-optimized baseline for each QNN. Thr.put (FPS) is throughput in frames per second.}\label{tab: end2end-overview-table}
\pgfplotstableread[col sep=tab,trim cells=true]{data/end2end_summary.tsv}\qnnresults
\small
\pgfplotstabletypeset[
columns={Network,Acc,Thr,LUT,rLUT,BRAM,rBRAM,DSP,rDSP,Thr.put (FPS),Latency (ms)},
display columns/0/.style={string type, column type=c},
display columns/1/.style={string type, column type=c},
display columns/2/.style={string type, column type=c},
display columns/3/.style={column type=r, fixed, dec sep align},
display columns/4/.style={column type=r},
display columns/5/.style={column type=r},
display columns/6/.style={column type=r},
display columns/7/.style={column type=r},
display columns/8/.style={column type=r},
display columns/9/.style={column type=r},
display columns/10/.style={column type=r},
display columns/11/.style={column type=r},
every head row/.style={before row=\toprule, after row=\midrule},
every last row/.style={after row=\bottomrule},
assign column name/.style={/pgfplots/table/column name={\textbf{#1}}}
]{\qnnresults}
\end{table}

\autoref{tab: end2end-overview-table} provides the FPGA resources and performance results for our QNN workloads, following the methodology described in \autoref{sec:methodology-end2end}. 
As \OurScheme{} optimizations are intended for resource savings, the degree of parallelization for each network stays constant across optimizations, and we do not see differences in throughput and latency.
The single exception is TFC-w2a2, which shows a noticeable latency reduction of 50\% when thresholding conversion is enabled.
Since TFC-w2a2 is a very small workload with a high degree of parallelization, a microsecond reduction in latency due to fewer kernel instantiations in the thresholding layer tails becomes noticeable.

\subsubsection{Effect of \OurScheme{} Optimizations.} 
\label{sec:results-optimization-breakdown}
\begin{figure}
    \centering
\begin{tikzpicture}
        \begin{groupplot}[ybar stacked,/pgf/bar width=5.0, group style={group size=4 by 3, xlabels at=edge bottom, ylabels at=edge left},height=3cm,width=4cm,ylabel={Resources}, legend style={at={(-1.6,-0.75)}, anchor=south}, legend columns=2, ymin=0,
        symbolic x coords={B, A, T, AT},
        xtick=data]
        \nextgroupplot[title={TFC-w2a2}, ylabel={LUT (k)}]
        \addplot+[ybar] plot coordinates {(B,28.76) (A,27.48) (T,28.76) (AT,27.48) };
        \addplot+[ybar] plot coordinates {(B,14.2) (A,13.64) (T,6.68) (AT,5.67) };

        \nextgroupplot[title={CNV-w2a2}]
        \addplot+[ybar] plot coordinates {(B,93.05) (A,92.87) (T,93.05) (AT,92.87) };
        \addplot+[ybar] plot coordinates {(B,31.85) (A,29.79) (T,27.09) (AT,25.28) };

        \nextgroupplot[title={RN8-w3a3}]
        \addplot+[ybar] plot coordinates {(B,222.73) (A,220.70) (T,222.73) (AT,220.76) };
        \addplot+[ybar] plot coordinates {(B,91.39) (A,83.54) (T,56.11) (AT,50.37) };

        \nextgroupplot[title={MNv1-w4a4}]
        \addplot+[ybar] plot coordinates {(B,33.26) (A,31.41) (T,33.26) (AT,31.41) };
        \addplot+[ybar] plot coordinates {(B,110.25) (A,106.02) (T,77.65) (AT,74.86) };

        \nextgroupplot[ylabel={BRAM}]
        \addplot+[ybar] plot coordinates {(B,0) (A,0) (T,0) (AT,0) };
        \addplot+[ybar] plot coordinates {(B,0) (A,0) (T,0) (AT,0) };

        \nextgroupplot[]
        \addplot+[ybar] plot coordinates {(B,148) (A,148) (T,148) (AT,148) };
        \addplot+[ybar] plot coordinates {(B,3.5) (A,3.5) (T,3.5) (AT,4) };

        \nextgroupplot[]
        \addplot+[ybar] plot coordinates {(B,324) (A,324) (T,324) (AT,324) };
        \addplot+[ybar] plot coordinates {(B,0) (A,0) (T,7) (AT,10) };

        \nextgroupplot[]
        \addplot+[ybar] plot coordinates {(B,481) (A,481) (T,481) (AT,481) };
        \addplot+[ybar] plot coordinates {(B,23) (A,23) (T,106.5) (AT,76) };

        \nextgroupplot[ylabel={DSP}]
        \addplot+[ybar] plot coordinates {(B,0) (A,0) (T,0) (AT,0) };
        \addplot+[ybar] plot coordinates {(B,13) (A,4) (T,5) (AT,0) };

        \nextgroupplot[]
        \addplot+[ybar] plot coordinates {(B,0) (A,0) (T,0) (AT,0) };
        \addplot+[ybar] plot coordinates {(B,22) (A,22) (T,5) (AT,0) };

        \nextgroupplot[]
        \addplot+[ybar] plot coordinates {(B,58) (A,58) (T,58) (AT,58) };
        \addplot+[ybar] plot coordinates {(B,64) (A,80) (T,5) (AT,0) };

        \nextgroupplot[]
        \addplot+[ybar] plot coordinates {(B,365) (A,365) (T,365) (AT,365) };
        \addplot+[ybar] plot coordinates {(B,57) (A,57) (T,5) (AT,0) };

        \addlegendentry{MAC layers}
        \addlegendentry{non-MAC layers}
        \end{groupplot}
    \end{tikzpicture}
    \caption{Breakdown of FPGA resources for MAC- and non-MAC layers. Labels refer to which optimizations were applied: (B)aseline, (A)ccumulator minimization, (T)hresholding, and (AT) for both.}
    \label{fig:end2end-breakdown}
\end{figure}
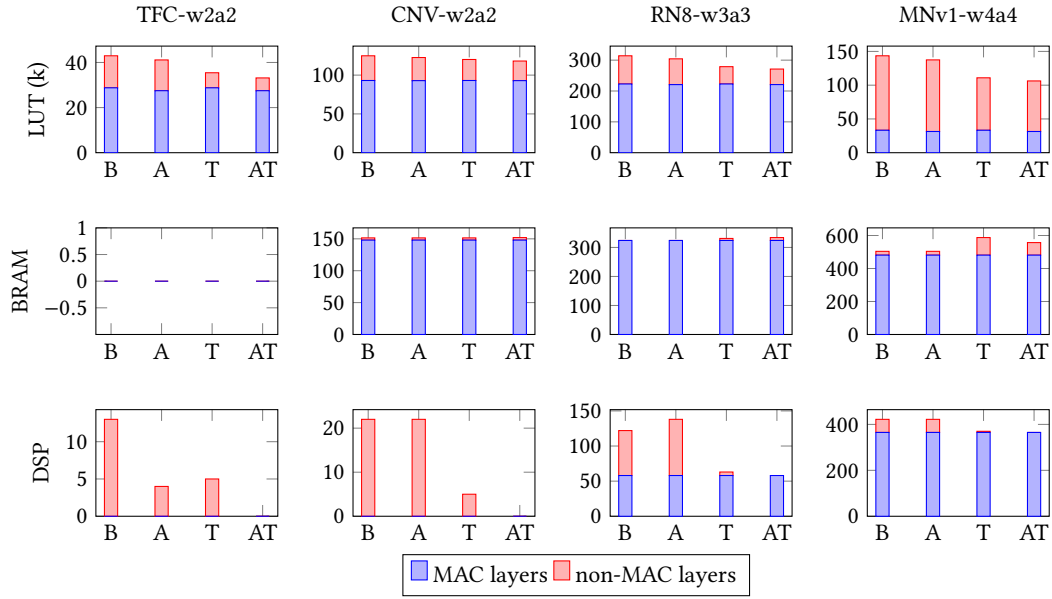

When both \OurScheme{} optimizations are enabled, we observe a 17\% reduction in LUTs and 66\% in DSPs, with a slight increase of 4\% in BRAMs.
Only enabling accumulator minimization yields a respective 3\% and 14\% reduction for LUTs and DSPs, while only enabling thresholding reduces LUTs and DSPs by 14\% and 50\% at a cost of increasing BRAMs by 5\%.
To better understand where the resource savings by \OurScheme{} optimizations originate, we take a closer look at the breakdown of FPGA resources for different layer types and optimizations. 
\autoref{fig:end2end-breakdown} provides a breakdown of the post-synthesis resources, including LUTs, BRAMs, and DSPs for MAC layers and non-MAC layers.
Here, the MAC layers category covers all resources associated with matrix multiplications originating from either fully-connected and convolutional layers, and non-MAC layers covers other components including FIFOs, data width converters, elementwise kernels, thresholding and others.

We first observe that across all networks and optimizations, the resources for the MAC layers are quite stable.
Thresholding does not affect MAC layer resources, and accumulator minimization yields an average LUT reduction of 3\% over the baseline.
Instead, the majority of savings from \OurScheme{} optimizations come from non-MAC layers, especially when accumulator width minimization and thresholding are combined.
The reduction in accumulator width propagates into downstream non-MAC layers and yields 6\% average LUT reduction on its own and 39\% when combined with thresholding.
In QNNs with higher precision, such as MNv1-w4a4, which use DSP computation for MACs, the majority of the LUT footprint is taken up by non-MAC operations, and the resource reductions with \OurScheme{} become more prevalent, with over 30\% reduction of LUTs achieved by enabling both accumulator minimization and thresholding.
The DSP resources themselves for non-MAC layers also see major reduction from \OurScheme{}-optimized layer tails, with 10\% reduction from accumulator minimization.
By combining accumulator minimization with thresholding, we note that DSPs consumed by non-MAC layers can be eliminated entirely, offering an attractive implementation alternative that retains the high accuracy of per-channel scales and biases without consuming valuable DSP slices that could be allocated for higher MAC layer throughput instead.
Finally, examining the cases where BRAMs are utilized, we note that the bulk of these resources are used for storing convolutional and fully-connected layer parameters and are largely unaffected by \OurScheme{} optimizations.
However, we notice an \emph{increase} in BRAMs for RN8-w3a3 (+3\%) and MNv1-w4a4 (+13\%) when thresholding is enabled, even in combination with accumulator minimization.
This hints at the expected trade-off for thresholding at higher activation bitwidths, which we characterize in more detail in \autoref{sec:results-microbenchmarks}.

\subsubsection{Minimized Accumulator Widths}
\label{sec:results-accwidth}

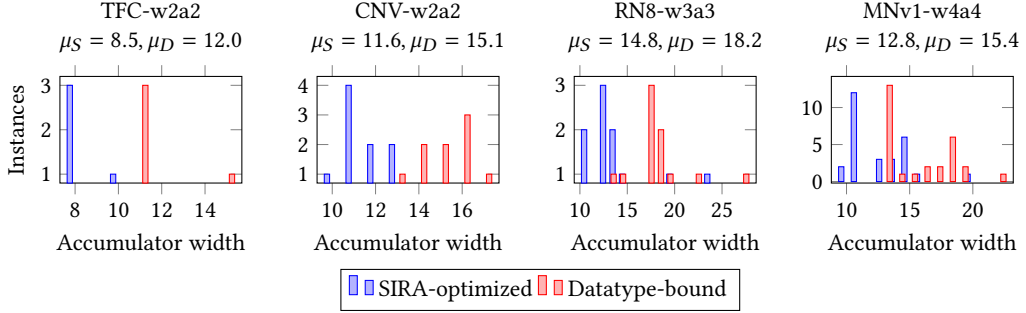
\begin{figure}
    \centering
    \begin{tikzpicture}
        \begin{groupplot}[ybar,/pgf/bar width=2.0, group style={group size=4 by 1, xlabels at=edge bottom, ylabels at=edge left,vertical sep=1.5cm},height=3cm,width=4cm,xlabel={Accumulator width},ylabel={Instances}, legend style={at={(-1.6,-1.2)}, anchor=south}, legend columns=2]
        \nextgroupplot[title style={align=center}, title={TFC-w2a2 \\ $\mu_{S} = 8.5, \mu_{D} = 12.0$}]
        \addplot table[col sep=tab]{data/acc-tfc-w2a2.tsv};
        \addplot table[col sep=tab]{data/dtb-tfc-w2a2.tsv};
        \nextgroupplot[title style={align=center}, title={CNV-w2a2 \\ $\mu_{S} = 11.6, \mu_{D} = 15.1$}]
        \addplot table[col sep=tab]{data/acc-cnv-w2a2.tsv};
        \addplot table[col sep=tab]{data/dtb-cnv-w2a2.tsv};
        \nextgroupplot[title style={align=center}, title={RN8-w3a3 \\ $\mu_{S} = 14.8, \mu_{D} = 18.2$}]
        \addplot table[col sep=tab]{data/acc-rn8-w3a3.tsv};
        \addplot table[col sep=tab]{data/dtb-rn8-w3a3.tsv};
        \nextgroupplot[title style={align=center}, title={MNv1-w4a4 \\ $\mu_{S} = 12.8, \mu_{D} = 15.4$}]
        \addplot table[col sep=tab]{data/acc-mnv1-w4a4.tsv};
        \addplot table[col sep=tab]{data/dtb-mnv1-w4a4.tsv};
        \addlegendentry{\OurScheme{}-optimized}
        \addlegendentry{Datatype-bound}
        \end{groupplot}
    \end{tikzpicture}
    \caption{Statistics for per-layer accumulator widths for the QNN workloads. $\mu_{S}, \mu_{D}$ are average accumulator widths optimized by \OurScheme{} and datatype-bound, respectively.}
    \label{fig:optimized-accwidths}
\end{figure}
\autoref{fig:optimized-accwidths} provides histograms for both datatype-bound and \OurScheme{}-optimized accumulator widths on a per-layer basis. 
We observe that the optimized accumulator width varies a great deal for our QNN benchmarks, varying from 8-bit for TFC-w2a2 to 24-bit for RB8-w3a3. 
As expected, we see that the optimized accumulator widths are primarily determined by the quantization for weights and activations, as well as the number of elements in each dot product. 
For instance, the 8-bit quantization used in the first and last layers of RN8-w3a3 and MNv1-w4a4 requires accumulators of 20 bits or more, in contrast to the rest of the layers.
For MNv1-w4a4, we additionally see a concentration of 11-bit accumulators. 
These are the accumulators for depthwise convolutions, which introduce dot products with fewer elements due to different channels being kept in separate dot products.
On average, accumulator widths optimized by \OurScheme{} are 63\% smaller compared to a 32-bit accumulation baseline, and 22\% compared to those calculated via the datatype bound.
How this translates into FPGA resource savings is mostly determined by the decrease in subsequent layer tail resources, as explored in \autoref{sec:results-optimization-breakdown}.

\subsection{Layer Tail Analysis}
\label{sec:results-microbenchmarks}

\subsubsection{Microbenchmarks}
\label{sec:result-microbenchmarks}
\newcommand{\mcc}[2]{\multicolumn{#1}{c}{\textbf{#2}}}
\newcommand{\mccv}[2]{\multicolumn{#1}{c|}{\textbf{#2}}}
\newcommand{\mcr}[2]{\multicolumn{#1}{r}{\textbf{#2}}}
\newcommand{\mrc}[2]{\multirow[c]{#1}{*}{\textbf{#2}}}
\newcommand{\best}[1]{{\color{green!35!black} \bf #1}}
\newcommand{\secondbest}[1]{{\color{orange!75!black} \bf #1}}
\newcommand{\worst}[1]{{\color{red!50!black} \bf #1}}
\begin{table}[!htbp]
    \caption{Layer Tail Microbenchmarks: Shown is LUT utilization of the entire layer tail. No other resources are utilized. 
    Marked in \best{green} is the best choice per category if only considering the resource utilization. In \worst{red} are instances where the multi-threshold representation uses even more resources than implementing composite float operations.}
    \small
    \begin{tabular}{ccrrrrrr|rrrrrr}
        \toprule
                      \mcr{2}{Granularity} & \mcc{3}{Per Tensor} & \mcc{3}{Per Channel} & \mcc{3}{Per Tensor} & \mcc{3}{Per Channel} \\
        \cmidrule(lr){3-5} \cmidrule(lr){6-8} \cmidrule(lr){9-11} \cmidrule(lr){12-14}
         & \mrc{1}{\textbf{Bits out}} & \mcc{1}{2} & \mcc{1}{4} & \mcc{1}{8} & \mcc{1}{2} & \mcc{1}{4} & \mcc{1}{8} & \mcc{1}{2} & \mcc{1}{4} & \mcc{1}{8} & \mcc{1}{2} & \mcc{1}{4} & \mcc{1}{8} \\
        \cmidrule{3-14}
        \addlinespace[2ex]
        \textbf{Scaling} & \textbf{Bits in}                  &                                                                  \mcc{6}{Thresholding} &                             \mcc{6}{Composite, \texttt{float32}} \\
        \mrc{3}{Free}         &  8 &  \best{119} &  \best{205} &   \best{430} &  \best{215} &  \best{715} &         7847  & 10541 & 10562 & 10580 & 10926 & 10944 & 10962 \\
                              & 16 &  \best{183} &  \best{310} &   \best{743} &  \best{375} & \best{1337} & \worst{16998} & 10633 & 10654 & 10672 & 11018 & 11036 & 11054 \\
                              & 24 &  \best{247} &  \best{419} &  \best{1044} &  \best{535} & \best{1947} & \worst{24991} & 10729 & 10750 & 10768 & 11114 & 11132 & 11150 \\

        \cmidrule{3-14}
        \mrc{3}{PoT}          &  8 &  \best{118} &  \best{185} &   \best{325} &  \best{209} &  \best{591} &         3560  & 10524 &  9469 & 10561 & 10722 & 10740 & 10758 \\
                              & 16 &  \best{182} &  \best{281} &   \best{485} &  \best{270} &  \best{695} &         4098  & 10616 &  9561 & 10653 & 10814 & 10832 & 10850 \\
                              & 24 &  \best{247} &  \best{383} &   \best{669} &  \best{338} &  \best{851} &         6568  & 10712 &  9657 & 10749 & 10910 & 10928 & 10946 \\
        \addlinespace[2ex]
        \cmidrule{3-14}

        \mcc{2}{}                  &                                                                  \mcc{6}{Composite, \texttt{fixed16.8}} &                          \mcc{6}{Composite, \texttt{fixed32.16}} \\
        \mrc{3}{Free}         &  8 &       2225  &       2280  &        2536  &       2344  &       2571  &   \best{3028} &  4387 &  4887 &  4909 &  4754 &  4914 &  5299 \\
                              & 16 &       2907  &       2974  &        3508  &       2734  &       3006  &   \best{3370} &  5185 &  5415 &  5840 &  5532 &  5484 &  6382 \\
                              & 24 &       3511  &       3595  &        4220  &       3284  &       3615  &   \best{4047} &  5590 &  5877 &  6265 &  5355 &  5302 &  6338 \\
        \cmidrule{3-14}
        \mrc{3}{PoT}          &  8 &       1820  &       1650  &        1880  &       2204  &       2224  &   \best{2264} &  3298 &  2841 &  3346 &  3721 &  3737 &  3769 \\
                              & 16 &       2544  &       2226  &        2593  &       2699  &       2715  &   \best{2748} &  4003 &  3598 &  4051 &  4637 &  4653 &  4685 \\
                              & 24 &       3073  &       2703  &        3119  &       3260  &       3278  &   \best{3306} &  4317 &  3881 &  4365 &  4569 &  4585 &  4617 \\
        \addlinespace[2ex]
        \bottomrule
        \addlinespace[2ex]
    \end{tabular}
    \label{tab:layer_tail_microbenchmarks}
\end{table}

\autoref{tab:layer_tail_microbenchmarks} shows the results of the layer tail benchmarks generated by synthesizing parameterizable instances of layer tails as shown in \autoref{fig:tail-modes}. 
As expected, for low bitwidth quantization up to 8 bits per-tensor or 4 bits per-channel, thresholding allows for the by far cheapest implementation of the layer tails in terms of LUT utilization, while at the same time guaranteeing numerical exactness. The other numerically exact implementation of layer tails in \texttt{float32} representation consistently utilizes more than an order of magnitude more LUTs and only becomes a viable options for larger bitwidths, where the growth of the thresholds memory becomes prohibitive. If trading numerical exactness for additional resource savings is acceptable, fixed-point representations offer a promising alternative for larger bitwidths, using only between 21\% and 58\% of the LUTs of the \texttt{float32} configurations.

Resource utilization of floating-point layer tails only marginally depends on the input and output bitwidth, which is expected as the output of a \texttt{float32} operation is always 32 bits, independent of the bitwidth of the input --- besides the first input and the final output, all parameter and intermediate tensors are in floating-point representation. For fixed-point layer tails, the requirements of lossless fixed-point arithmetic need to be accounted for as discussed in \autoref{sec:methodology_layer_tail_analysis}, hence the increased resource utilization with increasing bitwidths.
Both for thresholding and fixed-point representation, resource utilization clearly depends on the output bitwidth, as further analyzed in \autoref{sec:results-analytical-model}.

In general, finer scale, bias, and threshold granularity lead to increased resource utilization as expected, while restricting the scales to \ac{pot} enables further optimizations, resulting in reduced resource utilization even for the thresholding implementation. The latter effect hints at parameter value or distribution-dependent optimizations, which are currently not captured by our analytical models but are worth future investigations.

\subsubsection{Analytical Modeling} 
\label{sec:results-analytical-model}
\begin{figure}[htbp]
    \centering

    \begin{subfigure}[t]{0.45\textwidth}
    \centering
%

\definecolor{mycolor1}{rgb}{0.00000,0.44700,0.74100}%
\definecolor{mycolor2}{rgb}{0.85000,0.32500,0.09800}%
\definecolor{mycolor3}{rgb}{0.92900,0.69400,0.12500}%
\definecolor{mycolor4}{rgb}{0.49400,0.18400,0.55600}%
\definecolor{mycolor5}{rgb}{0.46600,0.67400,0.18800}%
\definecolor{mycolor6}{rgb}{0.30100,0.74500,0.93300}%
\definecolor{mycolor7}{rgb}{0.63500,0.07800,0.18400}%

\begin{tikzpicture}[%
trim axis left, trim axis right
]

\begin{axis}[%
height=0.65\columnwidth,
width=0.90\columnwidth,
xmin=1.00,
xlabel style={font=\color{white!15!black}},
xlabel={Output Bits},
xtick distance={1},
xlabel shift=-3pt,
ymin=0.00,
ymax=60000.00,
ylabel style={font=\color{white!15!black}},
ylabel={$\text{LUT}$},
axis background/.style={fill=white},
axis lines = left,
ymajorgrids,
yminorgrids,
ymode=log,
log ticks with fixed point,
legend style={legend cell align=left, align=left, draw=white!15!black},
legend style={at={(0.5,1.12)},anchor=center,font=\footnotesize},/pgf/number format/1000 sep={},
legend columns=2,
tick label style={/pgf/number format/assume math mode = true},
xlabel style={font=\normalsize},
ylabel style={font=\normalsize},
x tick label style={ /pgf/number format/.cd, fixed, fixed zerofill, precision=0, /tikz/.cd, font=\small},
y tick label style={ /pgf/number format/.cd, fixed, fixed zerofill, precision=0, /tikz/.cd, font=\small},
enlargelimits =false,clip=false,
samples=9,
domain=1:8
]

\addplot[mycolor1, only marks, domain=0:10, samples at={1,2,...,8}]{x * 24.0 * 4.0 + (2^x - 1)*24.0  * 32.0 / 64.0};
\addlegendentry{thres-chan-32}

\addplot[mycolor1, dashed]{(1.18 * 24*16 * 4 + 124)   + (2 *(24 +16 +16)* 4 + 24)    + (4.0 * (24+8+1) * 4 + 21 ) + (1.18 * 32*16 * 4 + 124) + ( 4.2 *(24+16+1)* 4 + 13 ) + 2*32*16 / 64 };
\addlegendentry{composite-chan-32}

\addplot[mycolor2, only marks, domain=0:10, samples at={1,2,...,8}]{x * 24.0 * 4.0 + (2^x - 1)*24.0  * 256.0 / 64.0};
\addlegendentry{thres-chan-256}

\addplot[mycolor2, dashed]{(1.18 * 24*16 * 4 + 124)   + (2 *(24 +16 +16)* 4 + 24)    + (4.0 * (24+8+1) * 4 + 21 ) +(1.18 * 32*16 * 4 + 124) + ( 4.2 *(24+16+1)* 4 + 13 ) + 2*256*16 / 64 };
\addlegendentry{composite-chan-256}

\addplot[mycolor3, only marks, domain=0:10, samples at={1,2,...,8}]{x * 24.0 * 4.0 + (2^x - 1)*24.0  * 512.0 / 64.0};
\addlegendentry{thres-chan-512}


\addplot[mycolor3, dashed]{(1.18 * 24*16 * 4 + 124)   + (2 *(24 +16 +16)* 4 + 24)    + (4.0 * (24+8+1) * 4 + 21 ) + (1.18 * 32*16 * 4 + 124) +( 4.2 *(24+16+1)* 4 + 13 ) + 2*512*16 / 64 };
\addlegendentry{composite-chan-512}
\end{axis}

\end{tikzpicture}
    \caption{Varying channels, fixed PE=4.}
     \label{fig:res-analytical-layertail-ch}
    \end{subfigure}
    \begin{subfigure}[t]{0.45\textwidth}
    \centering
%

\definecolor{mycolor1}{rgb}{0.00000,0.44700,0.74100}%
\definecolor{mycolor2}{rgb}{0.85000,0.32500,0.09800}%
\definecolor{mycolor3}{rgb}{0.92900,0.69400,0.12500}%
\definecolor{mycolor4}{rgb}{0.49400,0.18400,0.55600}%
\definecolor{mycolor5}{rgb}{0.46600,0.67400,0.18800}%
\definecolor{mycolor6}{rgb}{0.30100,0.74500,0.93300}%
\definecolor{mycolor7}{rgb}{0.63500,0.07800,0.18400}%

\begin{tikzpicture}[%
trim axis left, trim axis right
]

\begin{axis}[%
height=0.65\columnwidth,
width=0.90\columnwidth,
xmin=1.00,
xmax=8.0,
xlabel style={font=\color{white!15!black}},
xlabel={Output Bits},
xtick distance={1},
xlabel shift=-3pt,
ymin=0.00,
ymax=60000.00,
ylabel style={font=\color{white!15!black}},
ylabel={$\text{LUT}$},
axis background/.style={fill=white},
axis lines = left,
ymajorgrids,
yminorgrids,
ymode=log,
log basis y=10, 
log ticks with fixed point,
legend style={legend cell align=left, align=left, draw=white!15!black},
legend style={at={(0.5,1.12)},anchor=center,font=\footnotesize},/pgf/number format/1000 sep={},
legend columns=2,
tick label style={/pgf/number format/assume math mode = true},
xlabel style={font=\normalsize},
ylabel style={font=\normalsize},
x tick label style={ /pgf/number format/.cd, fixed, fixed zerofill, precision=0, /tikz/.cd, font=\small},
y tick label style={ /pgf/number format/.cd, fixed, fixed zerofill, precision=0, /tikz/.cd, font=\small},
enlargelimits =false,clip=false,
samples=9,
domain=1:8
]

\addplot[mycolor1, only marks, domain=0:10, samples at={1,2,...,8}]{x * 24.0 * 1.0 + (2^x - 1)*24.0  * 256.0 / 64.0};
\addlegendentry{thres-pe-1}

\addplot[mycolor1, thick, dashed]{(1.18 * 24*16 * 1 + 124)   + (2 *(24 +16 +16)* 1 + 24)    + (4.0 * (24+8+1) * 1 + 21 ) + (1.18 * 32*16 * 1 + 124) + ( 4.2 *(24+16+1)* 1 + 13 ) + 2*256*16 / 64 };
\addlegendentry{composite-pe-1}

\addplot[mycolor2, only marks, domain=0:10, samples at={1,2,...,8}]{x * 24.0 * 2.0 + (2^x - 1)*24.0  * 256.0 / 64.0};
\addlegendentry{thres-pe-2}

\addplot[mycolor2, thick, dashed]{(1.18 * 24*16 * 2 + 124)   + (2 *(24 +16 +16)* 2 + 24)    + (4.0 * (24+8+1) * 2 + 21 ) +  (1.18 * 32*16 * 2 + 124) + ( 4.2 *(24+16+1)* 2 + 13 ) + 2*256*16 / 64 };
\addlegendentry{composite-pe-2}

\addplot[mycolor3, only marks, domain=0:10, samples at={1,2,...,8}]{x * 24.0 * 4.0 + (2^x - 1)*24.0  * 256.0 / 64.0};
\addlegendentry{thres-pe-4}


\addplot[mycolor3, thick, dashed]{(1.18 * 24*16 * 4 + 124)   + (2 *(24 +16 +16)* 4 + 24)    + (4.0 * (24+8+1) * 4 + 21 ) + (1.18 * 32*16 * 4 + 124) +  ( 4.2 *(24+16+1)* 4 + 13 ) + 2*256*16 / 64};
\addlegendentry{composite-pe-4}

\end{axis}

\end{tikzpicture}
    \caption{Varying PE-parallelism, fixed channels=256.}
     \label{fig:res-analytical-layertail-pe}
    \end{subfigure}
    \caption{LUT cost prediction with thresholding and composite (\texttt{fixed16.8}) layer tails, both using 24-bit input, per-channel granularity.}
    \label{fig:thres_or_not}
\end{figure}

To characterize the cross-over points between composite and thresholding layer tail implementations, we apply the analytical models derived in \autoref{sec:analytical_cost_models} and plot the results in
\autoref{fig:thres_or_not}.
With the logarithmic y-axis, we see that the cost for manifest as straight lines due to the exponential growth with output bits, while the cost for composite tails are seemingly-constant levels with different offsets since their linear growth with output bits barely registers on the logarithmic scale.
The data indicate that for $<4$-bit outputs, thresholding is consistently the cheaper option, while composite tails are cheaper for $>8$-bit outputs.
Between 4 and 8-bit outputs, the tradeoff is less clear and depends on the number of channels and parallelism.
Because of the memory-dominated nature of thresholding cost, we see in \autoref{fig:res-analytical-layertail-ch} that a larger number of channels causes a notable increase.
For composite layer tails, which are compute-dominated, \autoref{fig:res-analytical-layertail-pe} shows a similar trend, where PE-parallelism shifts the level considerably.
Additionally, the cost of the composite layer tail is influenced by the input and parameter precision, whose variations we do not analyze here due to space limitations.

Overall, both our experiments in \autoref{tab:layer_tail_microbenchmarks} and the analytical model show clear winners for $<4$-bit (thresholding) and $>8$-bit (composite), whereas the 4-to-8-bit region needs more detailed consideration.
The choice of implementation style could be automated by integrating analytical models directly into \ac{fdna} compilers in future work, although the ability to manually select this per-layer remains valuable to address corner cases.

\section{Related Work}
\label{sec:related-work}
We present related work on range analysis, and how layer tails are handled in prior work on \acp{fdna}.

\subsection{Range Analysis for FPGA Circuits}
\label{sec:related-work-rangeanalysis}
A significant body of prior work has explored range and precision analysis for FPGA circuits in general, aiming to reduce resource usage by optimizing bitwidths.
Modern HLS tools commonly incorporate bitwidth-aware scheduling and binding, as proposed by
Cong et al.~\cite{cong2005bitwidth}, enhanced by bitmask analysis by Gort and Anderson \cite{gort2013range}.
Another field of study explores bitwidth optimization or word-length minimization, 
where the goal is to reduce the precision in a full-precision computational graph to balance area and error.
The work by Constantinides et al.~\cite{constantinides2003wordlength} notes that the underlying problem is NP-hard, and proposes solutions based on mixed integer linear programming or heuristics, while Lee et al.~\cite{lee2006accuracy} adopt analytical error models and simulated annealing.
The key difference in \OurScheme{} is that we do not aim to apply additional quantization to a computational graph. 
Instead, we treat neural network quantization as a separate problem which is already addressed by QAT or PTQ approaches (\autoref{sec:qnn-background}).
The computational graphs input to \OurScheme{} are fake-quantized graphs where integer computation is mixed with non-integer scales and biases, which we handle by tracking the components separately.
Although we do perform simple fixed-point quantization of aggregated scales and biases (\autoref{sec:fixed-pt-quant}) for non-fused implementations of layer tails, this is only to provide a reasonable baseline for comparing against thresholding-based layer tails and is not central to our approach.
At a finer granularity, as FINN uses Vitis HLS for implementing each layer individually, bitwidth-aware scheduling and binding optimizations~\cite{cong2005bitwidth} are already incorporated inside each layer, but the inter-layer bitwidth decisions are controlled by \OurScheme{}.
To the best of our knowledge, prior works do not address the problem of analyzing how scaled-integer intervals propagate in QNNs, and how this domain-specific knowledge can be utilized to optimize \acrshort{fdna}s.

\subsection{Layer tails in \acrshort{fdna} compilers}
\label{sec:handling-scale-factors}

We provide a brief overview of \acrshort{fdna} compilers, focusing on their treatment of layer tails arising from quantization, batch normalization, and other sources.

\subsubsection{Thresholding.}
\label{sec:related-finn-thresholding}
As described in Umuroglu~et~al.~\cite{umuroglu2017finn}, the original FINN compiler imported binarized neural networks, and absorbed full-precision scales and biases into thresholds of binary activation quantizers. This process of absorbing scale factors and biases into thresholds was termed \emph{streamlining} in \cite{umuroglu_jahre:CASES2017} and extended to multi-bit activations in Blott~et~al.~\cite{blott2018finn}. 
This is related to how \OurScheme{} propagates $\sis{v}$ and $\sib{v}$, but corresponds to a multi-step graph transformation instead of analysis and does not describe how to obtain the initial thresholds.
Extensions to FINN streamlining were developed for particular cases, enabling topologies such as MobileNet-v1 and ResNet-50~\cite{alonso2021elastic}. 
However, these works do not provide a concrete algorithm for streamlining, nor do they clarify restrictions that need to be applied to \acp{qnn} to make them streamlinable. 
Applying this form of streamlining to previously unseen quantizer and layer configurations is a trial-and-error process, which may require changing the quantization scheme or developing new graph transformations. 
Our work with \OurScheme{} provides clear guidelines on how to quantize neural networks in a way that makes scaled-integer propagation possible, and separates threshold conversion from scale and bias propagation to allow for more flexibility.

\subsubsection{Fixed-Point and Power-of-Two Quantization.}
The hls4ml framework~\cite{duarte2018fast} is an end-to-end tool that accepts input network description in Keras, PyTorch, and ONNX/QONNX frontends.
For each layer, HLS code is generated and compiled using Vivado/Vitis, Intel, or Catapult backends, depending on the target device.
A fixed-point format is used for implementing the scales and biases in the final deployment.
The work by Pappalardo et al.~\cite{qonnx_paper} briefly describes how scale factors are handled for fake-quantized graphs and mentions their propagation past linear operators for hls4ml. Handling of special cases, such as residual connections and depthwise convolutions, is not discussed.
Other \acrshort{fdna} compilers, such as FILM-QNN by Sun et al.~\cite{film-qnn} and HPIPE by Hall and Betz~\cite{hpipe}, fold \texttt{BatchNormalization} parameters into the preceding convolution layer, and apply fixed-point quantization for scales and biases.
NN2FPGA by \citet{nn2fpga} utilizes the uniform affine quantization scheme (\autoref{sec:qnn-background}) with quantization scales restricted to power-of-two values and a zero-point equal to $0$, also folding \texttt{BatchNormalization} into the preceding convolution layer. 
Scaling by power-of-two constraints is handled by bit shifts. %
However, folding batch normalization in this manner can lead to significant accuracy loss in low-precision models at 4 bits or fewer \cite{neseem2024pikelpn}, as well as requiring richer scale factors to maintain task performance with low-bit quantization, as discussed in \autoref{sec:qnn-background}.

\subsection{Layer tails in \acp{fdna}}
\label{sec:other_work}

To our knowledge, most prior work on \ac{fdna}s does not explore the alternatives for implementing the layer tail, opting for a particular combination of quantization scale, granularity, and hardware implementation that works for their use case.
For comparison with previous work from the perspective of how quantization scales and other high-precision operations like \texttt{BatchNormalization} are implemented, we discuss several prior \acp{fdna} for the commonly-used CIFAR-10 and ImageNet datasets.
The implementations are summarized in \autoref{tab:other_implementations}.
Most works implement a quantization scheme that allows utilizing bitshifts to scale the output activations of a layer before processing them by the next layer.
Wu et al. \cite{mnv2_19} utilize 8-bit fixed-point quantization for the MobileNet-v2 topology.
Lu et al. \cite{lu2024resnet} presented an FPGA implementation of ResNet-8 using 16-bit fixed-point quantization with hls4ml.
Hamanaka et al. \cite{hamanaka2023resnet} implemented 4-bit fixed-point quantization for a ResNet-8 topology using FINN.
Although the fixed-point quantization scheme would allow for implementing quantization by a simple bitshift, their work mentions that thresholding is used, which makes it likely that \texttt{BatchNormalization} layers were absorbed into thresholds.
Another approach is to follow the work of Jacob et al. \cite{jacob2018quantization}, where the scaling after each MAC operation is implemented by multiplication with a fixed-point approximation of the quantization scales, followed by a bitshift.
The work of Hong et al. presents an implementation of MobileNet-v1 using 8 bits for weights and activation.
The quantization scales of convolution weights, input, and output are aggregated into a single fixed-point parameter.
Yan et al. \cite{mnv2_21} also follow this approach and implement an 8-bit MobileNet-v2 model, where the \texttt{BatchNormalization} operation is fused with the quantization scales.

Overall, comparing accuracy scores for both prior works and our work supports the claim that more expressive scale factors better maintain accuracy, especially in the low-bit quantization regime.
With thresholding, our CNV-w2a2 achieves comparable accuracy to the 16-bit ResNet-8 by Lu et al.~\cite{lu2024resnet}, and our MNv1-w4a4 exhibits almost 2\% points better accuracy compared to the 8-bit MobileNet-v2 implementation by Wu et al.~\cite{mnv2_19}.

\begin{table}[]
\caption{Layer tail details and accuracy for prior \acp{fdna} on CIFAR-10 and ImageNet.}
\label{tab:other_implementations}
\small
\begin{tabular}{@{}llllllll@{}}
\toprule
Dataset  & Work                                                       & Topology     & Precision & Scale constraint          & Scale impl.\textsuperscript{+}               & \texttt{BatchNorm.}\textsuperscript{*} & Accuracy \\ \midrule
CIFAR-10 & This work & CNV     & w2a2      & float & thr              & thr       & 88.8     \\
CIFAR-10 & This work & CNV     & w2a2      & float & fix              & fix       & 87.9     \\
CIFAR-10 & Hamanaka et al. \cite{hamanaka2023resnet} & ResNet-8     & w4a4      & \ac{pot} & thr              & thr       & 85.9     \\
CIFAR-10 & Lu et al. \cite{lu2024resnet}             & ResNet-8     & w16a16    & float  & fix & sca       & 89.2     \\
CIFAR-10 & Bosio et al. \cite{nn2fpga}               & ResNet-8     & w4a4      & \ac{pot} & shi                  & con        & 86.9     \\
CIFAR-10 & Bosio et al. \cite{nn2fpga}               & ResNet-8     & w8a8      & \ac{pot} & shi                  & con        & 88.7     \\
\midrule
ImageNet &This work                 & MobileNet-v1 & w4a4      & float & thr                         & thr          & 69.9     \\
ImageNet &This work                 & MobileNet-v1 & w4a4      & float & fix                         & fix          & 68.5     \\
ImageNet & Wu et al. \cite{mnv2_19}                 & MobileNet-v2 & w8a8      & \ac{pot} & unk                         & fix          & 68.1     \\
ImageNet & Yan et al. \cite{mnv2_21}                & MobileNet-v2 & w8a8      & float  & fix & sca       & 70.8     \\
ImageNet & Hall and Betz \cite{hpipe}                & MobileNet-v2 & w16a16    & \ac{pot} & unk                         & con        & 71.7     \\
ImageNet & Hong et al. \cite{hong2024mnv1}           & MobileNet-v1 & w8a8      & float  & fix & unk               & 71.5     \\ 
\midrule
\multicolumn{8}{l}{\textsuperscript{+}Scale implemented as (fix)ed-point, by folding into (thr)esholds, (shi)fting, (unk)own } \\
\multicolumn{8}{l}{\textsuperscript{*}\texttt{BatchNormalization} implemented by folding into (thr)esholds, (sca)les, (con)volution, or as (fix)ed-point, (unk)own} \\
\bottomrule
\end{tabular}
\end{table}

\section{Conclusion}
\label{sec:conclusion}
As MAC precision decreases, non-MAC operations play an increasingly important role in total \ac{qnn} inference cost.
In this paper, we present \OurScheme{}, which employs interval arithmetic to extract integer range, scale, and bias information for each tensor in a trained QNN.
Based on this information, we show how various optimizations for \ac{fdna} compilation can be achieved, in particular by accumulator width minimization beyond the data type bound, aggregation of scales and biases, and conversion of groups of elementwise operations to thresholding layers for streamlined deployment.
Extending the FINN compiler with these optimizations, we demonstrated an average reduction of 17\% for LUTs, 66\% for DSPs, and 22\% for accumulator bitwidths with \OurScheme{}-powered optimizations for a range of QNNs for computer vision tasks.
To guide dbetween thresholding-based and composite implementations of QNN layer tails for \acrshort{fdna} compilers and designers, we offer microbenchmarks and analytical models, indicating that the bitwidth of quantized activations, the parallelism of the layer tail, and the granularity of scales and biases are the key determinants for efficiency.

Restricted range and precision via quantized weights and activations has already resulted in greatly reduced DNN inference cost, and we believe that further range-aware training and compilation of QNNs can bring insights to further increase efficiency.
Our results already reveal several interesting phenomena that warrant further study in numerical behavior in trained QNNs, including stuck channels (\autoref{sec:results_stuck_channels}), and we also see potential for incorporating prior work on automatic fixed-point quantization \cite{constantinides2003wordlength, lee2006accuracy}.
From the analysis side, relaxing constraints on static scales and biases would enable \OurScheme{} extending to more QNNs, including those using asymmetric quantization for weights.
Moving to more sophisticated range analysis algorithms beyond interval arithmetic may also offer further optimization opportunities via tighter bounds on ranges.
In terms of further \acrshort{fdna} optimizations, we believe thresholding has potential for more study and optimization, including opportunities for threshold compression and quantification of trade-offs for more sophisticated activation functions.
Last but not least, with the emergence of \acrshort{fdna} compilers supporting transformer-style architectures \cite{berganski2024finnt}, support beyond convolutional and MLP-style topologies is an important direction for broader applicability for \OurScheme{}.

\section*{Acknowledgements}

\copyright 2025 Advanced Micro Devices, Inc. 
All rights reserved.
AMD, the AMD Arrow logo, Vivado, Vitis, Kria, Kintex, UltraScale, and combinations thereof are trademarks of Advanced Micro Devices, Inc.
Other product names used in this publication are for identification purposes only and may be trademarks of their respective companies.

\bibliographystyle{ACM-Reference-Format}
\bibliography{defs,refs}
\end{document}